\documentclass[12pt,a4paper]{article}
\usepackage{latexsym}

\usepackage[dvips]{graphics,lscape,rotating}
\usepackage{euscript}
\usepackage{psfrag}
\usepackage{bbm}
\usepackage{amsfonts,amsmath,amssymb}
\usepackage{mathrsfs}
\usepackage{pstricks,array,longtable}

\numberwithin{equation}{section}

\newcommand{\lo}[1]{^{#1}\!\!}

\newcommand{\cmt}[2]{\mbox{\begin{minipage}[t]{#1cm} #2 \end{minipage}}}
\newcommand{\mb}[1]{\mathbbm{#1}}  
\newcommand{\Muserfunction}[1]{A}

\newcommand{\PI}[3]{\big[\pi_{{#2}}(#1)\big]_{{#3}}}
\newcommand{\PIbar}[3]{\overline{\big[\pi_{{#2}}(#1)\big]}_{{#3}}}
\newcommand{\PIinv}[3]{\big[\big(\pi_{{#2}}(#1)\big)^{-1}\big]_{{#3}}}

\newcommand{\Sket}[3]{\big|~#1~#2~#3~\big>}
\newcommand{\ket}[1]{\big|~#1~\big>}
\newcommand{\NN}{\nonumber\\}

\newcommand{\barr}[1]{\begin{eqnarray}\begin{array}{#1}}
\newcommand{\earr}{\end{array}\end{eqnarray}}

\newcommand{\WT}[1]{\widetilde{#1}}
\newcommand{\MC}[1]{\mathcal{#1}}

\newcommand{\mineval}{\lambda_{\hat{V}}^{(\mathrm{min})}}

\newcommand{\maxev}{\lambda_{\hat{V}}^{(\mathrm{max})}}
\newcommand{\maxeval}{\maxev}
\newcommand{\eval}{\lambda_{\hat V}}
\newcommand{\jmax}{j_{\mathrm{max}}}
\newcommand{\nbins}{N_{\mathrm{bins}}}
\newcommand{\nevals}{N_{\mathrm{evals}}}
\newcommand{\numevals}{N_{\mathrm{evals}}}
\newcommand{\sigconf}{$\vec{\sigma}$-configuration}
\newcommand{\sigconfs}{$\vec{\sigma}$-configurations}
\newcommand{\signconf}{$\vec{\epsilon}$-configuration}

\def\ba{\begin{eqnarray}}
\def\ea{\end{eqnarray}}
\def\be{\begin{equation}}
\def\ee{\end{equation}}

\newtheorem{Theorem}{Theorem}[section]

\DeclareMathOperator{\tr}{tr}

\begin{document}

\title{Properties of the Volume Operator\\ in Loop Quantum Gravity I: Results}

\author{Johannes Brunnemann\thanks{ brunnemann@math.uni-hamburg.de}\\
Department of Mathematics, University of Hamburg, 20146 Hamburg, Germany\\
\\
David Rideout\thanks{drideout@perimeterinstitute.ca}\\
Perimeter Institute for Theoretical Physics, Waterloo, Ontario N2L 2Y5, Canada}
\maketitle

\begin{abstract}
We analyze the spectral properties of the volume operator of  
Ashtekar and  
Lewandowski in Loop Quantum Gravity, which is the quantum analogue of the classical volume expression for regions in three dimensional Riemannian space. 
Our analysis considers for the first time generic graph vertices of valence greater than four. 
Here we find that the geometry of the underlying vertex characterizes the spectral properties of the volume operator, in particular the presence of a `volume gap' (a smallest non-zero eigenvalue in the spectrum) is found to depend on the  
vertex embedding.   
We compute the set of all non-spatially diffeomorphic non-coplanar vertex embeddings for vertices of valence 5--7, and 
argue that these sets can be used to label
spatial diffeomorphism invariant states. 
We observe  
how gauge invariance connects vertex geometry and representation properties of the underlying gauge group in a natural way.
Analytical results on the spectrum of 4-valent vertices are included,  
for which  the presence of a volume gap is shown. This paper presents our main results; details are provided by a companion paper \cite{NumVolSpec}.
\end{abstract}
\section{ Introduction}
Loop Quantum Gravity (LQG) \cite{TT:big script,Rovelli: LQG Intro,Ashtekar:2004eh} has become a promising candidate for a quantum theory of gravity over the last 15 years. It is an attempt to canonically quantize General Relativity (GR) while preserving its key principle: background independence\footnote{
With the term `background independence' we imply independence on a choice of fixed background geometry.  In order to allow for the possibility of topology change in quantum gravity, it would be  
desirable to have the theory in its final formulation be independent of the topology of the underlying manifold, however to our knowledge
this has not been achieved so far within LQG.}.  
The resulting quantum theory is formulated as an $SU(2)$ gauge theory.

Upon casting GR into a Hamiltonian formulation by introducing a foliation of four dimensional spacetime $(\MC{M},g_{\mu\nu})$\footnote{Here $g_{\mu\nu}$ denotes the four dimensional metric on $\MC{M}$ which 
is a solution to Einstein's equations. Note that in this context, because of the diffeomorphism invariance of General Relativity, the metric  
is a representative of an equivalence class  
of metrics which can be transformed into each other by  
spacetime diffeomorphisms.} 
into spatial three dimensional hypersurfaces $\Sigma$, with the  
orthogonal timelike direction parametrized by $t\in\mb{R}$, 
one obtains first class constraints  which have to be imposed on the reformulated theory such that it obeys the dynamics of Einstein's equations and is independent of the particular choice of foliation. These constraints are the three spatial diffeomorphism or vector constraints which generate diffeomorphisms inside $\Sigma$, and  
the so called Hamilton constraint which generates deformations of the hypersurfaces $\Sigma$ in the $t$- (foliation) direction. 
In addition one obtains three Gauss constraints due to the introduction of additional $SU(2)$ gauge degrees of freedom.
(There is also a more recent reformulation of the constraints, as proposed in the master constraint programme. See \cite{Thiemann:2005zg} and references therein.) 
The resulting theory is then quantized on the kinematical level in terms of holonomies $h$ and electric fluxes $E$. Kinematical states are defined over collection of edges of  
embedded graphs. The physical states have to be constructed by imposing the operator version of the constraints on the thus defined kinematical theory.

There exist well defined operators in the kinematical quantum theory which correspond to (classical) differential geometric objects, such as the length
of curves \cite{TT:Length Operator}, the area of surfaces \cite{AL: Quantized Geometry: Area Operators}, and the volume of regions \cite{Rovelli Smolin Volume Operator,AL: Quantized Geometry: Volume Operators}  in the spatial foliation hypersurfaces $\Sigma$. Remarkably all these quantities have discrete spectra, which can be traced back to the compactness of the gauge group $SU(2)$.

The thus constructed kinematical theory has already proved to potentially answer a number of outstanding physical questions.  Matter can be coupled to gravity without any problem, and the ultraviolet divergences which plague conventional quantum field theory (QFT) on a fixed background disappear if geometry and matter are simultaneously quantized \cite{TT:QSD V}. A coherent state framework \cite{TT: GCS I} has been formulated
which allows one to address questions on the correct semiclassical limit of the theory as well as to examine the possible effect of a dynamical background on the propagation of matter \cite{Sahlmann:2002qj,Sahlmann:2002qk}. Moreover, the entropy of a black hole has been calculated from first principles \cite{Rovelli:1996dv,Ashtekar:1997yu,Meissner:2004ju} in the LQG framework. 
\\

A central role in all these investigations is played by the area and the volume operator respectively. The former is understood very well analytically \cite{AL: Quantized Geometry: Area Operators} and numerically \cite{Helesfai Bene}  and enters the construction of coherent states \cite{Sahlmann:2002qj,Sahlmann:2002qk,Sahlmann: CS for CQG} and the computation of black hole entropy \cite{Rovelli:1996dv,Ashtekar:1997yu,Meissner:2004ju}. 
However, less is known about the latter, due to its complicated structure. Nevertheless it is a crucial object not only in order to analyze matter coupling to LQG, but also for evaluating the action of the Hamilton constraint (alternatively master constraint) operator in order to construct the physical sector of LQG. 
By construction the spectral properties of the constraint operators are driven by the spectrum of the volume operator.
In addition, the analysis of the classical big bang singularity \cite{triad I,triad II} heavily depends on the volume operator. 
As 
is obvious, controlling this operator is highly desirable, if one wants to truly test the present framework of LQG. In particular it is of utmost importance to check whether the promising results obtained in symmetry reduced frameworks (see e.g.\ \cite{Bojowald:2006ip,Bojowald:2006da,Husain:2003ry,Husain:2004yz} and references therein) remain valid in full unreduced LQG. 
From this perspective the present work is also intended to serve as a starting point for numerical studies which explore physical predictions from the full theory.

Historically two quantizations of the classical volume expression have been worked out:
In \cite{Loll 1} Loll has  constructed a quantized volume expression on a fixed lattice. 
This expression was later slightly changed \cite{Loll 2} and in \cite{Lewandowski:1996gk} recognized to be compatible with the construction by Ashtekar and Lewandowski $\hat{V}^{(AL)}$ in \cite{Ashtekar:1994wa} and later  
in \cite{AL: Quantized Geometry: Volume Operators}.
The second construction $\hat{V}^{(RS)}$ \cite{Rovelli Smolin Volume Operator} due to Rovelli and Smolin uses a different regularization scheme.
These two operators differ in their invariance properties: while the operator $\hat{V}^{(RS)}$ is invariant under homeomorphisms of the spatial foliation hypersurfaces $\Sigma$, the version $\hat{V}^{(AL)}$ contains `sign factors' which are invariant only under diffeomorphisms in $\Sigma$, its spectrum is sensitive to non-differentiable transformations.

The action of $\hat{V}^{(RS)}$ on the kinematical basis states has been analyzed in \cite{de Pietri II}, in \cite{de Pietri} its matrix elements are computed for the special case of four-valent vertices and in \cite{Seifert I,Seifert II} bounds on the eigenvalues for general four-valent vertices and monochromatic $n$-vertices are given. 
A closed formula for the general matrix elements of $\hat{V}^{(AL)}$, as well as eigenvalues for special spin configurations, have been obtained in \cite{TT:vopelm}. However, the derived expression suffers from its complicated structure which in practice prevents a more detailed analysis. The situation was significantly changed with the results of \cite{Volume Paper}, in which a drastically simplified expression for the matrix elements of $\hat{V}^{(AL)}$ was obtained.  This opens the way to analyzing $\hat{V}^{(AL)}$ in general as presented below.

In \cite{Consistency Check I,Consistency Check II} the two operator versions $\hat{V}^{(RS)}$, $\hat{V}^{(AL)}$ have been compared with respect to their consistency upon using an alternative electric flux quantization\footnote{The idea is to regard the volume operator as fundamental and to derive an operator corresponding to the flux operator from the volume. One may then check whether the thus constructed alternative flux quantization is in agreement with the usual one. See section \ref{Loop Quantum Gravity}.}. The result is that using  $\hat{V}^{(AL)}$ is consistent, while the usage of $\hat{V}^{(RS)}$ leads to problems. 
For this reason we employ the regularization used in the construction of $\hat{V}^{(AL)}$, and 
imply $\hat{V}^{(AL)}$ when referring to the volume operator in the remainder. 
\\

Of course, 
one may ask several questions at this point: Why  should we expect a quantum theory of gravity to be sensitive to the differentiable structure of the underlying (foliation) hypersurfaces? Shouldn't we regard this as a structure only to be re-obtained in a classical limit of the ultimate theory?  Also, why should we consider vertex valences greater than four?
One can introduce a triangulation of a three dimensional foliation hypersurface $\Sigma$ by decomposing it into tetrahedrons. This cell decomposition is then dual to a graph having four valent vertices only, one vertex inside each tetrahedron. As the volume operator acts only non-trivially at the vertices (see below), we can well interpret this as a cell decomposition of $\Sigma$, where each cell comes with a volume according to the action of the volume operator.
Moreover the Hamilton constraint operator as constructed in \cite{TT:QSD V} does not increase the valence of the vertices of a graph, so the four-valent sector seems to be preserved under `evolution' in foliation direction $t$.

We would like to take a rather conservative point of view here: By construction LQG is a minimalist theory in the sense that, starting as close as possible to General Relativity, one introduces as few additional structures as necessary in order to construct a quantum theory of gravity. One then wishes to analyze the fate of classical quantities in the quantum theory.  The quantization in terms of holonomies and fluxes seems to be a natural choice in order to quantize in a background independent way. We can construct a representation of the classical holonomy-flux-algebra on the kinematical Hilbert space, where holonomies act by multiplication, fluxes by differentiation, on states. Here it is important to note that 
multiplication is well defined for holonomies in an \emph{arbitrary} graph, which contains \emph{arbitrary} valent vertices.
Our intent is to study the predictions of the theory as it is stated formally, without making any assumptions regarding the nature of physical states.

Of course, the construction in terms of graphs suggests an interpretation of the kinematical theory in terms of a triangulation of $\Sigma$, hence it is tempting to directly associate graphs as parts of a triangulation.
However, one also might take the point of view that graphs are only abstract auxiliary constructions one must introduce in order to describe a meaningful quantum theory, but \emph{without} a direct interpretation. It would then be desirable to \emph{derive} a possible interpretation from the abstract Hilbert space formulation rather than introduce it by hand. In particular one should be able to show from first principles of the abstract formulation that it is sufficient to restrict to certain 
(e.g.\ 4-valent) types of graphs.
\\

With this motivation at hand we have numerically\footnote{The structure of the matrix representation of the volume operator becomes very complicated already for vertices of valences of 5 and above. This in general prevents an analytical treatment. See however section \ref{Analytical Results on the Gauge Invariant 4-Vertex} for analytical results on the 4-vertex.} analyzed the spectrum of the volume operator at non-coplanar vertices (no three edge tangents lie in a single plane)
of valences 4--7, and also the widely used \cite{Sahlmann:2002qk,ALQG} cubic\footnote{This vertex can be compared to the lattice points in a cubic lattice. Certainly opposite edges have parallel tangents in this case, hence there are triples of coplanar edge tangents.} 6-vertex. 
We find that the particular embedding of a graph into $\Sigma$ on the one hand and the recoupling of representations of the gauge group $SU(2)$ on the other hand are combined in the spectral properties of the volume operator. 
The distribution of eigenvalues at a given vertex is driven by the embedding of this vertex, expressed in terms of (spatial) diffeomorphism invariant sign factors which label the relative orientations of edges at the vertex.  In particular the existence of a volume gap depends on the vertex embedding. We find that eigenvalues accumulate close to zero in general. 
For the cubic six-valent vertex, however,  
we observe the presence of a volume gap.
Also, for arbitrary  
gauge invariant four-valent vertices, we  
show the existence
of a volume gap and give an analytic lower bound for the spectrum under certain mild assumptions on the edge spins.

The diffeomorphism invariant sign factors, moreover, allow for a classification of possible non-diffeomorphic embeddings of a vertex having $N$ edges, and can be used as a discrete label for diffeomorphism invariant states\footnote{As shown in \cite{Rovelli: Moduli Spaces}, there also exists a set of continuous diffeomorphism invariant parameters, called moduli, which additionally label a vertex. However the volume operator is not sensitive to this information, and classifying states with respect to their volume spectrum results in the described discrete set of vertex classifications labelled by sign factors.}. We will show how the possible embeddings of an $N$-valent vertex can be computed numerically and analytically.
As it turns out, there is an important interplay between gauge invariance and sign factors which leads to cancellations in the volume contributions for certain large valence vertices. This suggests that it may be possible to formulate conditions for which higher valent vertices can indeed be neglected. A more detailed analysis will be presented in a forthcoming paper \cite{VertexCombinatorics}.   
\\

This paper is organized as follows.  In section 2 we will  recall basic steps in the construction of LQG as necessary in order to explain the starting point for our analysis. At the same time this will introduce our notation. 
Note that this introductory chapter is not reproduced in the companion paper \cite{NumVolSpec}.
Experts may safely skip this section.
The construction of the volume operator $\hat{V}^{(AL)}$ is sketched in section 3, in particular the occurrence of the sign-factors is illustrated there. We sketch the implementation of the volume operator on a computer and discuss the problem of computing gauge invariant basis states as well as the set of possible non-diffeomorphic embeddings of vertices into the three dimensional foliation hypersurfaces $\Sigma$. 
Section 4 provides  the results of a detailed numerical spectral analysis of the volume operator for vertex valences $5,6,7$, where the sensitivity of $\hat{V}$ to the diffeomorphism invariant geometric properties of the underlying vertex is demonstrated.
Section 5 finally states analytic results on the volume operator action at four-valent vertices, in particular we  
give an analytic lower bound for the smallest non-zero eigenvalues. 
We close with a summary and outlook section 6 where we point to future application of our results, as well as the computational tools developed during this project.
\\

We  
close the introduction with a side remark.  The present publication summarizes our most important results. The interested reader is referred to \cite{NumVolSpec}  
to study
the details on each of the results presented here. We recommend reading this summary paper first in order to get an overview of what has been achieved, before studying  
the 
lengthy detailed presentation in \cite{NumVolSpec}.
    
\section{\label{Loop Quantum Gravity}Loop Quantum Gravity}
\subsection{Classical Theory}
\subsubsection{Hamiltonian Formulation}
In order to cast General Relativity into the Hamiltonian formalism, one has to perform a foliation $\MC{M}\cong\mb{R}\times\Sigma$ of the four dimensional spacetime manifold $(\MC{M},g_{\mu\nu})$ into three dimensional spacelike hypersurfaces $\Sigma$ with transverse time direction labelled by a foliation parameter $t\in\mb{R}$. Note that this foliation has to be seen as an auxiliary construction and it is kept completely arbitrary, since General Relativity is manifestly invariant under four dimensional diffeomorphisms of spacetime. 

As analyzed by Arnowitt, Deser and Misner \cite{ADM}, the four dimensional metric $g_{\mu\nu}$ can then 
be decomposed into the three metric $q_{ab}(x)$ on the spatial slices $\Sigma$\footnote{As induced by $g_{\mu\nu}$, that is the projection of $g_{\mu\nu}$ on the spatial slice.} and its extrinsic curvature $K_{ab}(y)$\footnote{That is the rate of change of $q_{ab}(x)$ with respect to $t$. This 
encapsulates the information of $g_{\mu\nu}$ in the foliation direction.},
which serve as canonical variables. ($x$ and $y$ are points in $\Sigma$.) However it seems to be difficult to formulate a quantum theory out of this approach.  These problems can be overcome if one introduces new variables due to Ashtekar \cite{Ashtekar Variables}: one may implement additional gauge degrees of freedom and take as canonical variables  electric fields $E^a_i(x)$ and connections $A_b^j(y)$ as used in the canonical formulation of Yang Mills theories. 
Here $a,b=1\ldots 3$ denote spatial (tensor) indices, $i,j=1\ldots 3$ denote $su(2)$-indices. 

The occurrence of $SU(2)$ as a gauge group comes from the fact that it is necessary for the coupling of spinorial matter and it is the universal covering group of $SO(3)$ which arises naturally when one rewrites the three metric $q_{ab}$ in terms of cotriads as 
\be\label{q in terms of e} 
   q_{ab}=e^j_ae^k_b\delta_{jk} .
\ee 
Obviously this definition is only unique up to $SO(3)$ rotations.  

The pair $(A,E)$ is related to $(q,K)$ as
\be\label{classical canonical variables}
   A^j_a=\Gamma^j_a-\beta K_a^j~~~~~~~~~~~~~E^a_j=\frac{1}{\beta}\sqrt{\det(q)}e^a_j
\ee
with $\Gamma$ being the spin connection (a certain function of the triads $e^j_a$ and its first spatial derivatives), $K^j_a$ is related to the extrinsic curvature $K_{ab}$ of the three metric by $K_{ab}=K^i_a e^j_b \delta_{ij}$ 
and $e^a_j$ is the inverse of $e^j_a$, that is $e^a_je^j_b=\delta_b^a$, $e^a_je^k_a=\delta_j^k$. Moreover we have the so called  Immirzi parameter $\beta$ \cite{Immirzi}, which in principle can be chosen as an arbitrary complex constant not specified a priori, but is chosen to be real in most cases\footnote{It is  possible to set $\beta$ to a complex number, which tremendously simplifies the form of the constraints below. However this simplification comes at the price of ensuring the reality of geometric quantities such as the spatial metric $q_{ab}$. It turns out that these reality conditions are very difficult to implement in the quantum theory. This is the reason why $\beta$ is usually chosen to be real. }. 
The pair $(A,E)$ obeys the Poisson bracket:
\be\label{class P A}
   \big\{E^a_i(x),A_b^j(y) \big\}=\kappa \delta^a_b\delta^j_i\delta(x,y)
\ee
with $\kappa=8\pi G_N$, $G_N$ being Newton's constant. All other Poisson brackets vanish. Note that (\ref{class P A}) is invariant under a rescaling of the Immirzi parameter $\beta$. For clarity of the notation we will suppress this $\beta$ dependence in the following and only insert it where necessary. 
\subsubsection{Constraints}

The constructed theory is subject to constraints which arise due to the background independence of General Relativity and the resulting non-bijective nature of the Legendre transformation performed upon introducing conjugated momenta. It can be treated as suggested by Dirac \cite{Dirac Lectures on QM}. One obtains the 
three vector (spatial diffeomorphism) constraints
\be\label{diffeo constraint}
   C_a=F^j_{ab}~E^b_j
\ee
where $F^j_{ab}=\partial_a A^j_b -\partial_b A^j_a +\epsilon_{jkl} A^k_aA^l_b$ is the curvature of the connection $A$. Moreover one gets a scalar(Hamilton) constraint
\be\label{Hamilton constraint}
   C=\frac{1}{\sqrt{|\det(E)|}}~\epsilon_{jkl}
               \Big[F^j_{ab} - (1+\beta^2)\epsilon_{jmn}K^m_aK^n_b \Big] E^a_kE^b_l
\ee
with $K^j_a=A^j_a-\Gamma^j_a(E)$. Additionally, due to the introduction of additional $SU(2)$ degrees of freedom in (\ref{q in terms of e}), one obtains the 
three Gauss constraints
\be\label{Gauss constraints}
   C_j=\partial_a~E^a_j+\epsilon_{jkl}A^k_aE^a_l 
\ee

\subsubsection{Regularization of the Poisson Algebra} 
We have shown how General Relativity can be rewritten as a theory with constraints in the Hamiltonian formalism. Moreover we have chosen a polarization of the classical phase space, that is we have chosen a split into a configuration space $\MC{A}$, consisting of smooth $SU(2)$-connections $A^j_b(x)$ defined on every point $x$ of the spatial foliation hypersurfaces $\Sigma$\footnote{In the following we will refer to an $A\in\MC{A}$ as a particular smooth configuration, given by its values $A^j_b(x)$ at every $x\in \Sigma$.}, and
a momentum space $T^*\MC{A}$, the cotangent bundle containing the momenta $E_i^a(x)$.
 
As mentioned above we must
enlarge the phase space by introducing $SU(2)$ gauge degrees of freedom, that is we can perform gauge local $SU(2)$ transformations $g\in \MC{G}$, where $\MC{G}$ denotes the space of smooth gauge transformations and every $g$ is defined locally as $g(x)$ for all $x\in\Sigma$.   
\\

Now the Poisson algebra (\ref{class P A}) has to be regularized in order to construct a meaningful quantum theory.
Therefore one uses the following smeared\footnote{For details on the regularization procedure we refer to \cite{TT:big script}.} quantities :
\begin{itemize}
  \item{The dual\footnote{With respect to spatial degrees of freedom.} $*E^a_j(x)$ (a 2-form) of $E^a_j(x)$ integrated over a two dimensional orientable piecewise analytic surface $S$, embedded in $\Sigma$ by\newline $\lo{(S)\,}X: \big[-\frac{1}{2},\frac{1}{2}\big]^2 \ni (u^1,u^2)\mapsto S=X(u^1,u^2)$~~:
  \be\label{smeared E}
     E_i(S)
     :=\int\limits_S *E_i=\int\limits_S d^2u~ \epsilon_{abc}\, \lo{(S)\,}X^b_{,u^1}(u^1,u^2)\,\lo{(S)\,}X^c_{,u^2}(u^1,u^2)E^a_i\big(\lo{(S)\,}X(u^1,u^2)\big)  \ee}
     
  \item{The integrated version of $A_b^j(y)$ (a 1-form) over a one dimensional oriented piecewise analytic\footnote{Edges should intersect only finitely many times.} edge $e$ embedded in $\Sigma$ by \newline $e: [0,1]\ni t\mapsto e(t)$, that is the holonomy $h_e(A)$ of $A_b^j(y)$ given by
  \ba\label{holonomy definiing dgl solution}
    h_e(A)
    &:=&\mathcal{P}\exp{\Big[\int_e A\Big]} 
    =\mb{1}_2 +\sum_{n=1}^{\infty}
           \int\limits_{0}^1 dt_1\int\limits_{t_1}^1 dt_2 \ldots \int\limits_{t_{n-1}}^1 dt_n
           ~~A(t_1)A(t_2)\ldots A(t_n)~~~~~~~~~   
  \ea}
\end{itemize}
Here $\mathcal{P}\exp{\cdot}$ denotes the path ordered exponential.
Both $E_i(S)$ and $h_e(A)$ are by construction invariant under reparametrizations of the surface $S\subset\Sigma$ and the edge $e\subset\Sigma$. When we use the term `edge' in the following we will actually imply an equivalence class of edges, which differ only by a reparametrization. Such an equivalence class is sometimes also referred to as a path. 

Note that an arbitrary element $g$ of the set $\MC{G}$ of  smooth gauge transformations gives a map\linebreak $g: \Sigma\ni x\rightarrow g(x)\in SU(2)$, that is to every point $x\in \Sigma$ an element $g(x)$ of the gauge group $SU(2)$ is assigned. As the connection $A$ transforms under $g$ as $A\rightarrow A^g:=\mbox{ad}_g A - dg g^{-1}$ the holonomy $h_e$ transforms under $g$ as :
\be\label{gauge behavior of holonomy}
       h_e^g(A):=h_e(A^g)=~g(b(e))~h_e(A)~g(f(e))^{-1}
\ee
with $b(e)=e(t=0)$, $f(e)=e(t=1)$.
Now assume that the edge $e$ intersects $S$ at the intersection point $e(t_{int})$ given by the curve parameter $t_{int}$.
For an arbitrary edge $e$ one can always obtain a splitting $e=e_1\circ e_2$  such that the edge $e$ is decomposed into two edges $e_1,e_2$. Accordingly we then have  $h_e=h_{e_1}\cdot h_{e_2}$ where $\cdot$ denotes group multiplication. Moreover we can change the orientation of $e_1,e_2$ such that their common starting point is  $e(t_{int})$, then $h_{e^{-1}}=(h_e)^{-1}$. Therefore without loss of generality one can restrict oneself to consider only edges outgoing from $e(t_{int})$. Then (\ref{class P A}) can be rewritten as:
\ba\label{class P A reg simp}
   \big\{E_i(S),h_e(A) \big\}
   &=&\frac{\kappa}{4}~\epsilon(e,S)~\tau_i~h_{e}
\ea
Here the sign factor $\epsilon(e,S)=0$ if $e,S$ do not intersect or if $e\subset S$. If $e,S$ intersect in a point given by the curve parameter $t_{int}$ then $\epsilon(e,S)=1$ if the normal of $S$ and the tangent vector $\dot{e}(t_{int})$ point to the same side of $S$ and $\epsilon(e,S)=-1$ if they point to different sides. 
In what follows we always will assume such an adaption of edges to the surface $S$ when acting with $E_i(S)$.

\subsubsection{Cylindrical Functions}
Let us now consider a graph $\gamma$ embedded in the spatial hypersurface $\Sigma$, that is a collection $E(\gamma):=\{e_1,\ldots,e_N\}$ of $|E(\gamma)|:=N$ edges $(e_1,\ldots,e_N)$, having common points  
at their endpoints only.
($N$ can be countably infinite.)
The set of these intersection points of edges is called the vertex set $V(\gamma)$ of the graph $\gamma$. Moreover let us denote by $\Gamma$ the set of all such graphs $\gamma\subset\Sigma$.  

The set of holonomies $\{h_e\}_{e\in E(\gamma)}$ then defines a map $p_\gamma$ from the space $\MC{A}$ of smooth connections $A$ to the $N$ copies of the gauge group $SU(2)$ via:
\ba\label{p gamma 1}
   p_\gamma~:~~~~\MC{A}&\rightarrow& SU(2)^{|E(\gamma)|}
   \NN
   A&\mapsto&\{h_e(A)\}_{e\in E(\gamma)}
\ea
A function $f~:~\MC{A}\rightarrow \mb{C}$ is then called cylindrical  with respect to a graph $\gamma$ if it can be written as $f\equiv f_\gamma\circ p_\gamma$ such that
\be
   f_\gamma\big(h_{e_1}(A),\ldots,h_{e_N}(A)\big)~~:~~SU(2)^{|E(\gamma)|}\rightarrow \mb{C}
\ee
where $f_\gamma$ depends on the  holonomies of the edges $e\in E(\gamma)$, each defining a group element on one of the $N$ copies\footnote{One copy for each $e\in E(\gamma)$.} of the group $SU(2)$.
By applying Leibnitz' rule in the evaluation of the Poisson bracket (\ref{class P A reg simp}) one obtains:

\be\label{Regulated PA and cyl}
   \big\{E_i(S),f_\gamma(\{h_e(A)\}_{e\in E(\gamma)})\big\}
   = \frac{\kappa}{4} \sum_{e \in E(\gamma)}\sum_{AB}\epsilon(e,S)~
     \big[\tau_i h_e \big]_{AB}\frac{\partial}{\partial [h_e]_{AB}}
                                               f_\gamma(\{h_e(A)\}_{e\in E(\gamma)})
\ee
Note that $\big[\tau_j h_e\big]_{AB}
\frac{\partial}{\partial(h_e)_{AB}}$ 
can be identified with the action of the right invariant vector field $\big(X_jf\big)(g):=\frac{d}{dt}f(e^{t\tau_j}g)|_{t=0}$ on $SU(2)$,
that is, the vector field constructed by the transport of the tangent space $T_{\mathbbm{1}}SU(2)$ at
the unit element $\mathbbm{1}_{SU(2)}$ of $SU(2)$ to all elements $g \in SU(2)$ via right-multiplication with $g$.

\subsection{Quantum Theory}
\subsubsection{Flux and Holonomy Operators}
Now formally the  Poisson bracket $\big\{E_j(S),h_e\big\}$ can be turned into $\mb{i}\hbar \big[\hat{E}_j(S),\hat{h}_e\big]$,
the commutator of the according operators multiplied by $\mb{i}\hbar$.
This gives us $\hat{E}_j(S)$ acting as a differentiation and
$\hat{h}_e$ being the multiplication operator 
on 
the space of cylindrical functions  $\text{CYL}$ of the form $f=f_\gamma\circ p_\gamma$.  We will often suppress the decorations on $\hat{h}_e$.

Note that the formulation in terms of holonomies (\ref{holonomy definiing dgl solution}) and fluxes allows for connections $A\in\overline{\MC{A}}$ contained in the space of distributional connections $\overline{\MC{A}}$. In other words: Due to the compactness of $SU(2)$ the cylindrical functions are bounded and hence to each cylindrical function we can associate its $sup$-norm. Then the closure $\overline{\text{CYL}}$ with respect to the $sup$-norm can be associated to  $\overline{\MC{A}}$, which can be seen as the set of arbitrarily discontinuous maps from the set of edges (paths) to the gauge group $SU(2)$.  Moreover the gauge behavior  (\ref{gauge behavior of holonomy}) allows for gauge transformations $g\in\overline{\MC{G}}$ contained in the set of distributional gauge transformations $\overline{\MC{G}}$. To summarize:  $A^i_a(x)$ and $g(x)$ may vary arbitrarily within an infinitesimal neighborhood $\MC{U}(x)$ of a point $x\in \Sigma$.\footnote{In contrast to the classical theory which considers only smooth and even differentiable quantities which can be described by their Taylor series, this means that we have assignments $\overline{\MC{A}}\ni A: x\rightarrow A^i_a(x)$ and $\overline{\MC{G}}\ni g: x\rightarrow g(x)$ for every point $x$ in $\Sigma$, with no further restrictions such as differentiability or smoothness.}  Moreover we will denote by $\overline{\MC{A}}\big|_\gamma\subset\overline{\MC{A}}$ the restriction of the connection configurations $A\in \overline{\MC{A}}$ to the one dimensional subset $\gamma\subset\Sigma$.
Then (\ref{Regulated PA and cyl}) gives:
\ba  \label{operators II}
  \hat{E}^\gamma_i(S)~ [{f}_\gamma]
  =\mb{i}\hbar\big\{E_i(S),f_\gamma(\{h_e(A)\}_{e\in E(\gamma)})\big\} 
  &=&\frac{\mb{i}}{4}~\ell_P^2 \sum_{e\in E(\gamma)}
			        \epsilon(e,S) X_e^i~~ f_\gamma
\ea 
where we have denoted the right invariant vector field on the copy of $SU(2)$ belonging to a particular edge $e\in E(\gamma)$ by $X^i_e$, again $i=1,2,3$ being an $su(2)$ index\footnote{Note that an $su(2)$-index can be shifted from upper to lower position (and vice versa) using the Cartan-Killing-metric $\delta_{ij}$. However since $\delta_{ij}=\delta^{ij}=\mb{1}_{ij}$ is the unit matrix, the index position makes no difference. Hence we will adjust this position as necessary for notational convenience.}.
Here naturally the Planck area shows up by $\ell_P^2=\hbar \kappa$.
Furthermore we pinned an index $\gamma$ to $E$ to denote that this operator acts on a function $f_\gamma$ restricted to the graph $\gamma$. One is then led to consider the Hilbert space $\MC{H}_{\gamma}:=L_2(\overline{\MC{A}}\big|_\gamma,d\mu_\gamma)$. $\MC{H}_{\gamma}$ is equipped with a natural inner product constructed from the measure $d\mu_\gamma:=\prod_{e\in E(\gamma)}d\mu_H(h_e)$ (the product of the Haar measures $d\mu_H$ on the individual copies of $SU(2)$ on the edges $e\in E(\gamma)$). It contains (the completion with respect to the $sup$-norm of) the    
set of cylindrical functions  $f_\gamma\big(h_{e_1}(A),\ldots,h_{e_N}(A)\big)$, $A\in\overline{\MC{A}}\big|_\gamma$, square integrable with respect to $d\mu_\gamma$.

\subsubsection{\label{SNF as Basis}Choosing a Basis in $\MC{H}_\gamma$: Spin Network Functions}
Recall that irreducible representations of  
$SU(2)$ can be realized by unitary  square matrices $\PI{h}{j}{}$ acting on a $(2j+1)$-dimensional linear vector space.
The label $j$  is a non-negative half integer ($j=0,\frac{1}{2},1,\ldots$)
and is called the weight of the representation $\pi_j$. 
Moreover matrix elements  $\PI{h}{j}{mn}$ of $\PI{h}{j}{}$ are usually labelled by $m,n=j,\ldots,-j$ from the upper left to the lower right corner.

Now due to the $Peter~\&~Weyl$ theorem the matrix element functions $\PI{\cdot}{j}{mn}$ of representation matrices of  $SU(2)$ are, up to a normalization factor, orthogonal with respect to the Haar measure $d\mu_H$ on $SU(2)$:    

\be\label{Peter & Weyl 1}
   \int_{SU(2)}d\mu_H(h) \PIbar{h}{j'}{m'n'} \PI{h}{j}{mn} 
   =\frac{1}{2j+1}\delta_{j'j}\delta_{m'm}\delta_{n'n}
\ee
Moreover they serve as an orthonormal basis on the space $L_2\big(SU(2),d\mu_H\big)$ of functions $f(h)$, $h\in SU(2)$,  square integrable with respect to the Haar measure $d\mu_H$ on $SU(2)$.
Using the multilabels $\Big\{\vec{j}:=(j_1,\ldots,j_N)$, $\vec{m}:=(m_1,\ldots,m_N)$ $\vec{n}:=(n_1,\ldots,n_N)\Big\}$  one can then define the so called Spin Network Functions (SNF) $T_{\gamma\vec{j}\vec{m}\vec{n}}$ for a fixed graph $\gamma$ as 
\be\label{Definition SNB}
   T_{\gamma\vec{j}\vec{m}\vec{n}}(A|_\gamma):=\prod_{e\in E(\gamma)}\sqrt{2j_e+1}\PI{h(A|_\gamma)}{j_e}{m_e n_e}
\ee
which serve then as an orthonormal basis on the Hilbert space $\mathcal{H}_\gamma=L_2(\bar{\mathcal{A}}|_\gamma,d\mu_\gamma)$ by applying (\ref{Peter & Weyl 1}) for every copy of $SU(2)$ belonging to the $N$ edges $(e_1,\ldots,e_N)$ in the edge set $E(\gamma)$ of the graph $\gamma$. 
Since every $L_2$ function $f=f_\gamma(h_{e_1},\ldots, h_{e_N})\circ p_\gamma$ cylindrical with respect to a graph $\gamma$  can be expanded in terms of the spin network basis (\ref{Definition SNB}) as
\be
   f_\gamma(A|_\gamma)=\sum_{\vec{j}~\vec{m}~\vec{n}} z_{(\vec{j}~\vec{m}~\vec{n})}~T_{\gamma~\vec{j}~\vec{m}~\vec{n} }(A|_\gamma)
\ee
with expansion coefficients $z_{(\vec{j}~\vec{m}~\vec{n})}$,
it suffices to work with the basis states $T_{\gamma\vec{j}\vec{m}\vec{n}}(A|_\gamma)$ in $\mathcal{H}_\gamma=L_2(\bar{\mathcal{A}}|_\gamma,d\mu_\gamma)$.

The inner product in $\mathcal{H}_\gamma$ can be extended to an inner product between SNFs $T_{\gamma\vec{j}\vec{m}\vec{n}}, T'_{\gamma'\vec{j}'\vec{m}'\vec{n}'}$ belonging to different fixed graphs $\gamma,\gamma'$ by choosing a bigger graph $\gamma''$ containing $\gamma,\gamma'$ as proper subsets, e.g.\ $\gamma''=\gamma\cup\gamma'$.\footnote{Of course one has to be careful in cases where edges of $\gamma,\gamma'$ intersect. Then one has to introduce additional vertices at these intersection points and to subdivide the intersecting edges, but this extension is straightforward.}  Then one may rewrite the SNFs $T_{\gamma\vec{j}\vec{m}\vec{n}}, T'_{\gamma'\vec{j}'\vec{m}'\vec{n}'}$ to depend on the additional edges according to the trivial representation, for example we set  $j_{e}=0$ in the rewritten $T_{\gamma''\vec{j}\vec{m}\vec{n}}$ of $T_{\gamma\vec{j}\vec{m}\vec{n}}$ if $e\in \{E(\gamma'')\setminus E(\gamma)\}$.

In order to construct a Hilbert space $\MC{H}_{kin}$ consisting of
cylindrical functions over {\it arbitrary} graphs from the subspaces
$\MC{H}_\gamma$ one has to take into account certain equivalence relations
between SNF's formulated for different graphs, that is $T_{\gamma}\sim
T'_{\gamma'}$, if the one can be obtained from the other by a subdivision of
edges (introducing trivial two valent vertices, see below) or a change of the
orientation of edges.  (The idea is to introduce a recoupling scheme for each
fixed graph $\gamma$ already at the gauge variant level and to exclude
trivial intertwiners at two valent vertices, see \cite{AL: Quantized
Geometry: Area Operators}, Sec IV.B for details.  As we will work at the gauge
invariant level we demonstrate a similar construction below in section
\ref{Gauge Invariance}.)

\subsubsection{Kinematical Hilbert Space $\MC{H}_{kin}$ of Loop Quantum Gravity}

The full kinematical Hilbert space  $\MC{H}_{kin}:=L_2(\overline{\MC{A}},d\mu_0)$ of Loop Quantum Gravity for the complete uncountable set of all graphs (which can be embedded in $\Sigma$) is then formulated as an inductive limit Hilbert space $\MC{H}_{kin}$ of directed sets of Hilbert spaces  
$\ldots\subset\MC{H}_\gamma \subset \MC{H}_{\gamma'}\subset \MC{H}_{\gamma''}\subset \ldots$ formulated over (infinite) directed sets of graphs  $\ldots\subset\gamma\subset\gamma'\subset\gamma''\subset\ldots$, where the direction goes to finer and finer graphs and the equivalence relations above are respected. 

$\MC{H}_{kin}$ can then be thought of as containing every possible such (infinite) directed set. Using (\ref{p gamma 1}) the measure $d\mu_0$ on $\MC{H}_{kin}$, called the Ashtekar-Lewandowski measure, is constructed such that for the measure $\mu_0(f)$ of a function $f\in\MC{H}_{kin}$ with $f=f_\gamma\circ p_\gamma$ one has:
\be
   \mu_0(f):=\int  f d\mu_0 = \int (f_\gamma \circ p_\gamma) d\mu_0
                          = \int f_\gamma  \big((p_{\gamma})_{\star} d\mu_0\big)
                          = \int f_\gamma  d\mu_\gamma    
\ee
that is $d\mu_\gamma$ can be written as a push forward $(p_{\gamma})_{\star}~ d\mu_0$ and we have the map $p_\gamma:~\MC{H}_{kin}\rightarrow \MC{H}_\gamma$.

The inductive limit construction of $\MC{H}_{kin}$ comes with the benefit
that in actual computations one can restrict oneself to the finite
dimensional graph level $\MC{H}_{\gamma}$, as long as one works with
operators which do not modify the graph ({\it cylindrical consistency}).  This
is the case for operators consisting of $\hat{E}^\gamma_i(S)$ only, such as
the area or volume operator. Then the results obtained on the level of a
single graph can be extended to $\MC{H}_{kin}$ in an obvious way. In what
follows we will restrict ourselves to the level of a single graph, that is
the finite dimensional Hilbert space $\MC{H}_{\gamma}$.

\subsubsection{\label{Gauge Invariance}Gauge Invariance} 
As can be seen in \cite{TT:big script}, the Gauss constraints $C_j$ as defined in (\ref{Gauss constraints}) form an ideal of the constraint algebra. Therefore they can be solved separately from the other constraints (\ref{diffeo constraint}), (\ref{Hamilton constraint}) by constructing gauge invariant spin network basis states.  

Let us start with a spin network basis state $T_{\gamma\vec{j}\vec{m}\vec{n}}(A)$ as defined in 
(\ref{Definition SNB}).
Using the possibility of redirecting  and splitting edges in the edge set $E(\gamma)$ of the underlying graph $\gamma$ mentioned in subsection \ref{SNF as Basis}, as well as the representation properties of the edge holonomies\footnote{That is $\PI{h_{e_1}h_{e_2}}{j}{mn}=\sum_{r} \PI{h_{e_1}}{j}{mr}\PI{h_{e_2}}{j}{rn}$ and 
$\PI{h_e^{-1}}{j}{mn}=\big[\big(\pi_j(h_e)\big)^{-1}\big]_{mn}$.},
we can 
introduce a new edge set $E(\widetilde{\gamma})$, where each vertex $v$ in the vertex set $V(\widetilde{\gamma})$ has only outgoing edges. Then we can rewrite $T_{\gamma\vec{j}\vec{m}\vec{n}}(A)$ as
\be\label{gauge Tgamma 0}
   T_{\gamma\vec{j}\vec{m}\vec{n}}(A)
   :=\left[\left(\bigotimes_{v\in V(\widetilde{\gamma})}\bigotimes_{\widetilde{e}\in E_v(\widetilde{\gamma})}
   \sqrt{2j_{\widetilde{e}}+1}\PI{h(A)}{j_{\widetilde{e}}}{}\right)\cdot M_{E(\gamma)}\right]_{\vec{m}\vec{n}}
\ee
where we take the tensor product now over the set of edges $E_v(\widetilde{\gamma})$ outgoing at each vertex $v\in V(\widetilde{\gamma})$, and the matrix $M_{E(\gamma)}$ contracts all representation matrices $\PI{h(A)}{j_{\widetilde{e}_1}}{}, \PI{h(A)}{j_{\widetilde{e}_2}}{}$ of two edges $\widetilde{e}_1,\widetilde{e}_2$ if their composition gives an edge $e$ of the original edge set $E(\gamma)$.
(For example $\widetilde{e}_1\circ (\widetilde{e}_2)^{-1}=e$, which implies
$\PI{h_e(A)}{j_e}{mn}=\sum_r \PI{h_{\widetilde{e}_1}(A)}{j_{\widetilde{e}_1}}{mr}
                     \PIinv{h_{\widetilde{e}_2}(A)}{j_{\widetilde{e}_2}}{rn}$.)
Moreover, in such cases, $j_{\WT{e}_1}=j_{\WT{e}_2}=j_e$.
In this way we have introduced trivial two-valent vertices $v_{e}^{(triv)}$ on each edge $e\in E(\gamma)$, which do not change the gauge behavior of the original function $T_{\gamma\vec{j}\vec{m}\vec{n}}(A)$ under a local gauge transformation $g\in \overline{\mathcal{G}}$, due to the transformation behavior (\ref{gauge behavior of holonomy}) of the holonomy.

Now by construction all edges in $E(\WT{\gamma})$ start at a vertex and end
at a trivial vertex $v_e^{(triv)}$, at which the gauge transformation
$g(v_e^{(triv)})$ does not have any effect by construction.  (Since at every trivial vertex $v_e^{(triv)}$ the representation $\PI{g(v_e^{(triv)})}{j_e}{}$ of the local gauge transformation $g(v_e^{(triv)})$ is multiplied by its inverse due to (\ref{gauge behavior of holonomy}).)
It follows that 
\be\label{gauge Tgamma 1}
   T_{\gamma\vec{j}\vec{m}\vec{n}}(A^g)
   :=\left[\left(\bigotimes_{v\in V(\widetilde{\gamma})}\bigotimes_{\widetilde{e}\in E_v(\widetilde{\gamma})}
   \sqrt{2j_{\widetilde{e}}+1}\PI{g(v)}{j_{\widetilde{e}}}{}\cdot\PI{h(A)}{j_{\widetilde{e}}}{}\right)\cdot M_{E(\gamma)}\right]_{\vec{m}\vec{n}}
\ee
Now for four matrices $(A,B),(C,D)$, each pair consisting of square matrices of the same dimension, we have 
$(A\cdot B)\otimes (C\cdot D)=(A\otimes C)\cdot (B\otimes D)$.  We can then rewrite (\ref{gauge Tgamma 1}) as \be\label{gauge Tgamma 2}
   T_{\gamma\vec{j}\vec{m}\vec{n}}(A^g)
   :=\left[\left(\bigotimes_{v\in V(\widetilde{\gamma})}
   \Big[\bigotimes_{\widetilde{e}\in E_v(\widetilde{\gamma})}
               \PI{g(v)}{j_{\widetilde{e}}}{}\Big]
   \cdot
   \Big[\bigotimes_{\widetilde{e}\in E_v(\widetilde{\gamma})}
              \sqrt{2j_{\widetilde{e}}+1}\cdot\PI{h(A)}{j_{\widetilde{e}}}{}\Big]\right)\cdot M_{E(\gamma)}\right]_{\vec{m}\vec{n}}
\ee  
It is obvious that we can obtain gauge invariant states $T_{\gamma\vec{j}\vec{I}}(A)$ if we use the fact that the tensor product of two irreducible representations  $\pi_{j_1}, \pi_{j_2}$ of $SU(2)$ can be reduced to a direct sum $\bigotimes_{j_{12}=|j_1-j_2|}^{j_1+j_2} \pi_{j_{12}}$ of irreducible representations $\pi_{j_{12}}$ whose matrix elements are constructed by using certain (anti-)symmetrizations of tensor product matrix elements \cite{NumVolSpec,Sexl Urbantke}. Therefore we can reduce the tensor product of representation matrix element functions at each vertex into a direct sum. Choosing a particular reduction order we can start with
\be\label{gauge Tgamma 3}
   \bigotimes_{\WT{e_1},\WT{e_2},\ldots,\WT{e}_{N_v}\in E_v(\widetilde{\gamma})}
              \PI{\cdot}{j_{\widetilde{e}}}{}
   =\bigoplus_{a_2(j_{\WT{e}_1}\,j_{\WT{e}_2} )=|j_{\WT{e}_1}-j_{\WT{e}_2}|} ^{j_{\WT{e}_1}+j_{\WT{e}_2}}
   \PI{\cdot}{a_2}{}\otimes \bigg[\bigotimes_{\WT{e_3},\ldots,\WT{e}_{N_v}\in E_v(\widetilde{\gamma})}                                    \PI{\cdot}{j_{\widetilde{e}}}{}
   \bigg]
\ee
where we denote the valence of a particular vertex $v$ by $N_v$.
By successive reduction the whole tensor product of representation matrix elements can be decomposed into a direct sum     
\be\label{gauge Tgamma 4}
   \bigotimes_{\WT{e_1},\WT{e_2},\ldots,\WT{e}_{N_v}\in E_v(\widetilde{\gamma})}
              \PI{\cdot}{j_{\widetilde{e}}}{}
   =\bigoplus_{a_2(j_{\WT{e}_1}\,j_{\WT{e}_2} )}
    \bigoplus_{a_3(a_2\,j_{\WT{e}_3} )}
    \cdots
    \bigoplus_{J(a_{N_v\!-\!1}\,j_{\WT{e}_{N_v}} )} 
      \PI{\cdot}{J}{}        
\ee 
The trivial representations $(J=0)$ contained in (\ref{gauge Tgamma 4}) then provide gauge invariant states, that is we can introduce {\bf intertwiners} 
$\vec{I}=\{I_v\}_{v\in V(\gamma)}$ which project onto the trivial representations at each vertex $v \in V(\WT{\gamma})$ which are contained in the decomposition (\ref{gauge Tgamma 4}):
\be\label{gauge Tgamma 5}
   T_{\gamma\vec{j}\vec{I}}(A)
   :=\left[\left(\bigotimes_{v\in V(\widetilde{\gamma})}
   I_v\cdot \Big[\bigotimes_{\widetilde{e}\in E_v(\widetilde{\gamma})} 
   \sqrt{2j_{\widetilde{e}}+1}\PI{h(A)}{j_{\widetilde{e}}}{}\Big]\right)\cdot M_{E(\gamma)}\right]
\ee
To summarize: Taking certain (anti-)symmetrizations of spin network functions  $T_{\gamma\vec{j}\vec{m}\vec{n}}(A)$, we can formulate at each vertex $v\in V(\gamma)$ intertwiners projecting onto the trivial subspaces contained in the \mbox{(anti-)} symmetrizations. Each particular set $\vec{I}=\{I_v\}_{v\in V(\gamma)}$ of projections then defines a gauge invariant state  $T_{\gamma\vec{j}\vec{I}}(A)$. 
  
It can be shown that the $T_{\gamma\vec{j}\vec{I}}(A)$ provide an orthonormal basis of the gauge invariant Hilbert space $\mathcal{H}=L^2(\overline{\mathcal{A}}/\overline{\mathcal{G}},d\mu_0)$, that is
\be\label{gauge Tgamma 6}
   \big<~T_{\gamma'\vec{j}'\vec{I}'}~,~T_{\gamma\vec{j}\vec{I}}~\big>
   ~=~
   \delta_{\gamma'\gamma}~\delta_{\vec{j}'\vec{j}}~\delta_{\vec{I}'\vec{I}}  
\ee
Note that the measure $d\mu_0$ is already gauge invariant, due to the translation invariance of the Haar measure $d\mu_H$ on $SU(2)$.

When gauge invariance is imposed, SNF's which differ by trivial two valent vertices are automatically identified: Consider a graph $\gamma$ consisting only of one closed loop $e$, its beginning/end point being the only vertex of $\gamma$.
It comes with a SNF $T_{\gamma j I}(A):=\tr_j\PI{h_{e}(A)}{j}{}$.
We now subdivide  $e$  into two segments $e=e_1\circ e_2$, introducing a
new graph $\WT{\gamma}$ which has a trivial two-valent vertex $v^{(triv)}_e$. We can then formally introduce a set of SNFs $T_{\WT{\gamma} \vec{j} \vec{m}\vec{n}}(A)
:=\PI{h_{e_1}(A)}{j}{m_1n_1} \PI{h_{e_2}(A)}{j}{m_2 n_2} $. 
However gauge invariance enforces 
$T_{\WT{\gamma} \vec{j} \vec{I}}(A):=\tr_j\PI{h_{e_1}(A)\cdot h_{e_2}(A)}{j}{}\equiv T_{\gamma j I}(A)$. 

\subsubsection{Spin States}

As it turns out, the action (\ref{operators II}) of the right invariant vector fields $X_e^k$ on spin network functions $T_{\gamma\vec{j}\vec{m}\vec{n}}$ can be related to the action of angular momentum operators $J_e^k$ as used in ordinary quantum mechanics  
by \cite{Consistency Check I,Consistency Check II,NumVolSpec}:
\be\label{correspondance to angular momentum}
   -\frac{\mb{i}}{2} X_e^k \PI{\cdot}{j_e}{m_e n_e}\equiv (-1)^{j_e+m_e}~J_e^k~ \Sket{j_e}{\!-\!m_e}{;n_e}
\ee
where $\Sket{j_e}{-\!m_e}{;n_e}$ denotes a usual angular momentum state
$\Sket{j_e}{-\!m_e}{\!\!\!}$ with an additional quantum number $n_e$ which is not affected by the action of the angular momentum operator $J_e^k$.

This opens the possibility of reformulating the theory using techniques from recoupling theory of angular momenta. In particular we may replace the tensor products at the vertices in (\ref{gauge Tgamma 5})
\be\label{introduce spin states}
   \Big[\bigotimes_{\widetilde{e}\in E_v(\widetilde{\gamma})} 
   \sqrt{2j_{\widetilde{e}}+1}\PI{\cdot}{j_{\widetilde{e}}}{}\Big]_{\WT{\vec{m}}_v\WT{\vec{n}}_v}
   =:\prod_{\tilde{e}\in E_v(\widetilde{\gamma})}\sqrt{2j_{\widetilde{e}}+1}\PI{\cdot}{j_{\widetilde{e}}}{\WT{m}_{\WT{e}}\WT{n}_{\WT{e}}}
  \equiv
  \prod_{\tilde{e}\in E_v(\widetilde{\gamma})}\Sket{j_{\widetilde{e}}}{\WT{m}_{\WT{e}}}{;\WT{n}_{\WT{e}}}
\ee
We may express the gauge invariant intertwiner expressions in terms of so called recoupling scheme standard basis states\footnote{See the appendix of the companion paper \cite{NumVolSpec} for details and definitions.}
$\big|\vec{a}~J\!=\!0~M\!=\!0~;~\vec{j}~\WT{\vec{n}}_v\big>_{(v)}$. 
Each of these states corresponds to a particular trivial $(J=0)$  representation contained in the reduction 
(\ref{gauge Tgamma 4}). Fix a labeling of the $N_v$ edges contained in the
set $\big\{\tilde{e}_1,\ldots,\tilde e_{N_{v}}\big\}=E_v(\WT{\gamma})$.  Each
of the trivial representations contained in (\ref{gauge Tgamma 4}) is then
uniquely labelled by the list of weights $a_2, a_3,\ldots,a_{N_v\!-\!1}$ of
the intermediate reduction steps in (\ref{gauge Tgamma 4}):\footnote{
The decoration `!' in `$\stackrel{!}{=}$' below simply indicates
that the equality is required to hold.}
\begin{footnotesize}
\ba\label{RC standard basis}
I_v\cdot \Big[\bigotimes_{\widetilde{e}\in E_v(\widetilde{\gamma})} 
   \!\!\sqrt{2j_{\widetilde{e}}+1}\PI{h(A)}{j_{\widetilde{e}}}{}\Big] 
   &\equiv&
   \big|a_2(j_1,j_2)~a_3(a_2, j_3)~\ldots a_{N_v\!-\!1}(a_{N_v\!-\!2}, j_{N_v\!-\!1})~
   J(a_{N_v\!-\!1}, j_{N_v})\!\stackrel{!}{=}\!0~~M\!\stackrel{!}{=}\!0~;~\vec{j}_v~\WT{\vec{n}}_v~\big>
   \NN
   &=:&\big|\vec{a}~J\!=\!0~M\!=\!0~;~\vec{j}_v~\WT{\vec{n}}_v\big>_{(v)}
\ea
\end{footnotesize}
Therefore 
\ba\label{Final definition of gauge inv snf}
  T_{\gamma\vec{j}\vec{I}}(\cdot)
   \equiv\left(\bigotimes_{v\in V(\widetilde{\gamma})}
   \big|\vec{a}~J\!=\!0~M\!=\!0~;~\vec{j}_v~\WT{\vec{n}}_v\big>_{(v)}\right)
   \cdot M_{E(\gamma)}
\ea

\section{The Volume Operator}
\subsection{Definition of the Volume Operator}

The operator corresponding to the volume $V(R)$ of a spatial region $R\subset\Sigma$
\ba\label{Classical Volume Expression}
   V(R)&=&\int_R~d^3x~\sqrt{\det q(x)}
   =\int_R~d^3x~\sqrt{\Big|\frac{1}{3!} \epsilon^{ijk}\epsilon_{abc}E^a_i(x)E^b_j(x)E^c_k(x) \Big|}
\ea
acting on gauge invariant\footnote{We restrict ourselves to the gauge invariant sector in the following. Almost everything could easily be formulated in terms of gauge variant spin network states.} spin network states, as derived in \cite{TT:vopelm,NumVolSpec} is defined as 
\ba
  \hat{V}(R)_{\gamma}~T_{\gamma\vec{j}\vec{I}}(\cdot)&=&\int\limits_R d^3p\widehat{\sqrt{det(q(p))_{\gamma}}}~T_{\gamma\vec{j}\vec{I}}(\cdot)
                     =\int\limits_R d^3p~ \hat{V}(p)_{\gamma}~T_{\gamma\vec{j}\vec{I}}(\cdot) 
\ea
where
\ba
  \label{nullte} \hat{V}(p)_{\gamma}&=&\ell_P^3 \sum_{v\in V(\gamma)}\delta^3(p,v)~\hat{V}_{v,\gamma}\\
  \label{erste} \hat{V}_{v,\gamma} &=&\sqrt{\Big| Z 
                   \sum_{I<J<K\le N_v} 
		   \epsilon (I J K)~ \hat{q}_{IJK}\Big|}\\
\label{definition qIJK}
   \hat{q}_{IJK}&=&\Big[(J_{IJ})^2,(J_{JK})^2\Big]   
\ea
and $(J_{IJ})^2=\sum_{k=1}^3 (J_I^k+J_J^k)^2$.  The construction of
(\ref{nullte}), (\ref{erste}), (\ref{definition qIJK}) is explained in detail
in \cite{NumVolSpec}.  Here $\ell_P$ denotes the Planck length, and $Z$ is a
real constant whose specific value was found in \cite{Consistency Check I} as
$Z=\beta^3\frac{3!\mb{i}}{4}\frac{1}{48}$.  For our purposes we set $Z$ to 1
or $\mb{i}$, since the Immirzi parameter $\beta$ can be freely chosen and $Z$
will only give an overall (at every vertex identical) constant scaling.  The
choice of $Z=1$ vs.\ $\mb{i}$ has the effect of rotating in the complex plane
the spectrum of the operator $Q$, defined in (\ref{Q}) below, from the
imaginary axis ($Z=1$) onto the real axis ($Z=\mb{i}$), which clearly makes
no difference.

The sum in (\ref{nullte}) has to be taken over all vertices $v\in V(\gamma)$ (each vertex having $N_v$ outgoing edges) of the underlying graph $\gamma.$ At each vertex $v$ one has to sum over all possible ordered triples $(e_I,e_J,e_K)\equiv (IJK)$, $I<J<K\le N_v$ of outgoing edges at  $v$. Here  $\epsilon(IJK)=\mbox{sgn}\big(\det{(\dot{e}_I(v),\dot{e}_J(v),\dot{e}_K(v))}\big)$ denotes the sign of the determinant of the tangents of the three edges $e_I,e_J,e_K$ intersecting at $v$. 

As is obvious from (\ref{nullte}), the volume operator acts nontrivially only at the vertices of the graph $\gamma$. Moreover its action at distinct vertices is completely unrelated\footnote{Though of course the vertices are related through the linking structure of the underlying graph $\gamma$.}. We can therefore restrict ourselves to analyze (\ref{erste}) at single gauge invariant vertices, since every graph $\gamma$ and hence every $T_{\gamma\vec{j}\vec{I}}$ can be constructed from those as can be seen from (\ref{Final definition of gauge inv snf}). {\it In what follows we will therefore consider only a single vertex $v$ and hence the single vertex operator $\hat{V}:=\hat{V}_{v,\gamma}$. Moreover we will denote the valence of the vertex by $N$, dropping the label $v$.}
  
\subsection{Computational Task}  
\subsubsection{Matrix Formulation}
The remaining task is then to consider
\ba\label{Volume definition}
   \hat{V}&:=&\sqrt{\Big|Z\cdot \sum\limits_{I<J<K\le N}\epsilon(IJK)~\hat{q}_{IJK} \Big|}
          =:\Big|Z\cdot Q \Big|^\frac{1}{2}
	  =\Big||Z|^2\cdot Q^\dagger Q \Big|^\frac{1}{4}
\ea
where $Q$ is a totally antisymmetric matrix with purely real elements. Its eigenvalues $\lambda_Q$ are purely imaginary and come in pairs $\lambda_Q=\pm\mb{i}\cdot\lambda$, $\lambda\ge 0$, $\lambda\in\mb{R}$.\footnote{$Q^\dagger Q$ is a totally symmetric real matrix, whose eigenvalues $\lambda_{Q^\dagger Q}$ are purely real and come in pairs $\lambda_{Q^\dagger Q}=\big|\lambda_Q\big|^2$.  This provides an alternate approach to computing the volume eigenvalues, which is discussed in \cite{NumVolSpec}.}
We choose $Z=1$.
Equation (\ref{Volume definition}) 
then tells us that the volume operator $\hat{V}$ has the same eigenstates as $Q$ 
 and its eigenvalues $\lambda_{\hat{V}}=\big|\lambda_Q\big|^\frac{1}{2}$.
Thus we are left with the task of calculating the spectra of totally antisymmetric real matrices of the form:
\be
   Q:=\sum\limits_{I<J<K\le N}\epsilon(IJK)~\hat{q}_{IJK}
   \label{Q}
\ee

\subsubsection{\label{Recoupling Computation}Computational Realization of the Recoupling Scheme Basis}
Here we briefly comment on the computation  
of the recoupling basis. 
In some cases the number of gauge invariant states contained in all possible recouplings of a set of spins can be surprisingly small. 
We are not aware of a
general exact analytic treatment of this problem, 
only
the case where all spins are taken to be $j=\frac{1}{2}$ \cite{ODreyer PC}.

A standard recoupling scheme is the successive recoupling of all $N$ angular momenta $j_1,\ldots,j_N$ at the vertex $v$, as follows.  Fix a labeling of the edges. Then recouple $j_1,j_2$ to a resulting angular momentum $a_2(j_1~j_2)$. We have, due to the Clebsch-Gordan-Theorem, $|j_1-j_2|\le a_2(j_1~j_2) \le j_1+j_2$.
Now couple $a_2,j_3$ to a resulting angular momentum $a_3(a_2~j_3)$ where again $|a_2-j_3|\le a_3(a_2~j_3)\le a_2+j_3$.  This continues until we arrive at the last recoupling step where $J\stackrel{!}{=} a_N(a_{N-1}~j_N)$ and thus $|a_{N-1}-j_N|\le J \le a_{N-1}+j_N$. If we work with gauge invariant recoupling schemes we have $J=0$ and therefore we must have $j_N\stackrel{!}{=}a_{N-1}$. 

The challenge is now to find, for a given spin configuration $j_1,\ldots,j_N$, all possible gauge invariant states, that is all recoupling sequences for which $a_k^{(min)}:=|a_{k-1}-j_k|\le a_k(a_{k-1}~j_k)\le (a_{k-1}+j_k)=:a_k^{(max)}$ (with $a_1=j_1$) and additionally $a_{N-1}\stackrel{!}{=}j_N$ is fulfilled \cite{NumVolSpec}.

\subsubsection{\label{Gauge Invariance at the Operator Level}Gauge Invariance at the Operator Level}
At the operator level gauge invariance  
\be
  (J_{(tot)})^2:=\sum_{k=1}^3\Big(\sum_{K=1}^N J_K^k\Big)^2=0
\ee
implies that 
\be\label{gauge invariance condition 2}
   J^k_N\stackrel{!}{=}-\sum_{L=1}^{N-1}J^k_L~~~\forall k=1,2,3
\ee
This can be used in (\ref{Volume definition}) in order to eliminate all $\hat{q}_{IJN}$ containing $N$ as an edge label \cite{NumVolSpec}.
Hence the volume operator can be rewritten as

\ba\label{Volume definition gauge invariant 3}
   \hat{V}&=:& \sqrt{\Big|Z\cdot 
          \sum\limits_{I<J<K<N}\sigma(IJK)~\hat{q}_{IJK} \Big|}        
\ea
where
\be\label{sigmas}
\sigma(IJK):= \epsilon(IJK)-\epsilon(IJN)+\epsilon(IKN)-\epsilon(JKN)
\ee
This is the final expression for the volume operator acting on gauge invariant vertex recoupling states. Expression (\ref{Volume definition gauge invariant 3}) generalizes the observation for the special case of a 4-valent vertex as stated in \cite{AL: Quantized Geometry: Volume Operators}, section V.D, and \cite{Volume Paper}.

\subsection{Implementation}
Obviously the analysis of (\ref{Volume definition gauge invariant 3}) contains two parts: a recoupling computation of the matrices $\hat{q}_{IJK}$ and a combinatorial analysis of the possible sign configurations $\sigma(IJK)$.

\subsubsection{\label{The constituent matrices}The $\hat{q}_{IJK}$ constituent matrices}
For a given spin configuration $j_1,\ldots,j_N$ at an $N-$valent vertex $v$ the trivial representations contained in the tensor product decomposition (\ref{gauge Tgamma 4}), that is the gauge invariant recoupling standard basis $\big|\vec{a}~J\!=\!0~M\!=\!0~;~\vec{n}~ \big>$ as given in (\ref{RC standard basis}), can be computed according to section \ref{Recoupling Computation}. A simplified general expression for the matrix elements of $\hat{q}_{IJK}$ was derived in \cite{Volume Paper} and will not be reproduced here.  An even more detailed discussion\footnote{There exist certain special edge combinations which have to be worked out carefully.} of this expression necessary for the present analysis is presented in \cite{NumVolSpec}. 
   
As we have stated before, an arbitrary but fixed labeling of the edges has to be chosen in order to perform the reduction (\ref{gauge Tgamma 4}). Since a relabeling of the edges can be implemented as a unitary basis transformation among the recoupling states (\ref{RC standard basis}), the spectrum of the volume operator is invariant under edge permutations.  (See \cite{NumVolSpec} for a more detailed discussion.)
We can use this property to drastically decrease the number of spin configurations to consider in the numerical analysis, because we can always choose a particular edge labeling $e_1,\ldots,e_N$ in which the spins are sorted 
\be
   j_1\le j_2\le \ldots \le j_N=j_{max}
\ee
Each of these spin configurations has degeneracy ${\bf{D}}(\vec{j})$ as follows.  Assume we have among the $N$ spins $\vec{j}=(j_1,\ldots,j_N)$ $p$ subsets of identical spins, each such subset containing $N_k$  ($k=1\ldots p$) elements. Such a configuration then stands for 
\be\label{Degenracy of the spin levels}
   {\bf{D}}(\vec{j})=\frac{N!}{N_1!\cdot N_2!\cdot\ldots\cdot N_p!}   
\ee
spin configurations.
Moreover we have the conditions $\sum_{n=1}^{N-1}j_n\ge j_N$ and $\sum_{n=1}^{N}j_n\in \mb{N}$ which are necessary in order to obtain a trivial representation in the reduction (\ref{gauge Tgamma 4}).

\subsubsection{Sign Configurations}

In \cite{Rovelli: Moduli Spaces} it was shown that for vertices of valence greater than 4, which are formed by self intersecting\footnote{These are essentially embeddings of $S^1\rightarrow \Sigma$.} smooth edges, there exist continuous moduli parameters labeling the thus formed knot in a diffeomorphism invariant way.
However there is also a discrete set of information  
regarding the linear dependence of edge tangents contained in the tangent space at the intersection point: Three tangents can be coplanar or linearly independent. Moreover, as we are going to show now, there is diffeomorphism invariant information on the relative orientation (right handed / left handed) of linearly independent triples of tangents: If we have five or more edges at a vertex, then we can look at the orientation of each triple of tangents at that vertex. We find that (up to an overall orientation flip) only special combinations of tangent orientations can be embedded in $\Sigma$.

Let us   
denote by $\vec{\epsilon}$ the set of signs $\epsilon(IJK)=0,\pm 1$, one for each ordered edge triple $(e_I,e_J,e_K)$ with $I<J<K\le N$.  
We have ${N\choose 3}$ such signs describing a particular embedding of an $N$-valent vertex $v$.
According to this definition we denote by $\vec{\sigma}$ the set of values
$\sigma(IJK)=0,\pm 1,\pm 2,\pm 3,\pm 4 $ computed from a particular set $\vec{\epsilon}$ via (\ref{sigmas}) for each of the 
${N-1 \choose 3}$ ordered edge triples $(e_I,e_J,e_K)$ with $I<J<K< N$.
Note that the map $\MC{S}:~\vec{\epsilon}\mapsto\vec{\sigma}$ is non-injective, in that there can be more than one $\vec{\epsilon}$-configuration giving the same $\vec{\sigma}$-configuration. We will denote this multiplicity by $\chi_{\vec{\sigma}}:=|\{\mbox{sign configurations~} \vec{\epsilon}\mbox{~for which}: \MC{S}(\vec{\epsilon})=\vec{\sigma}\}|$.
\\

\begin{figure}[htbp]
\center
\includegraphics{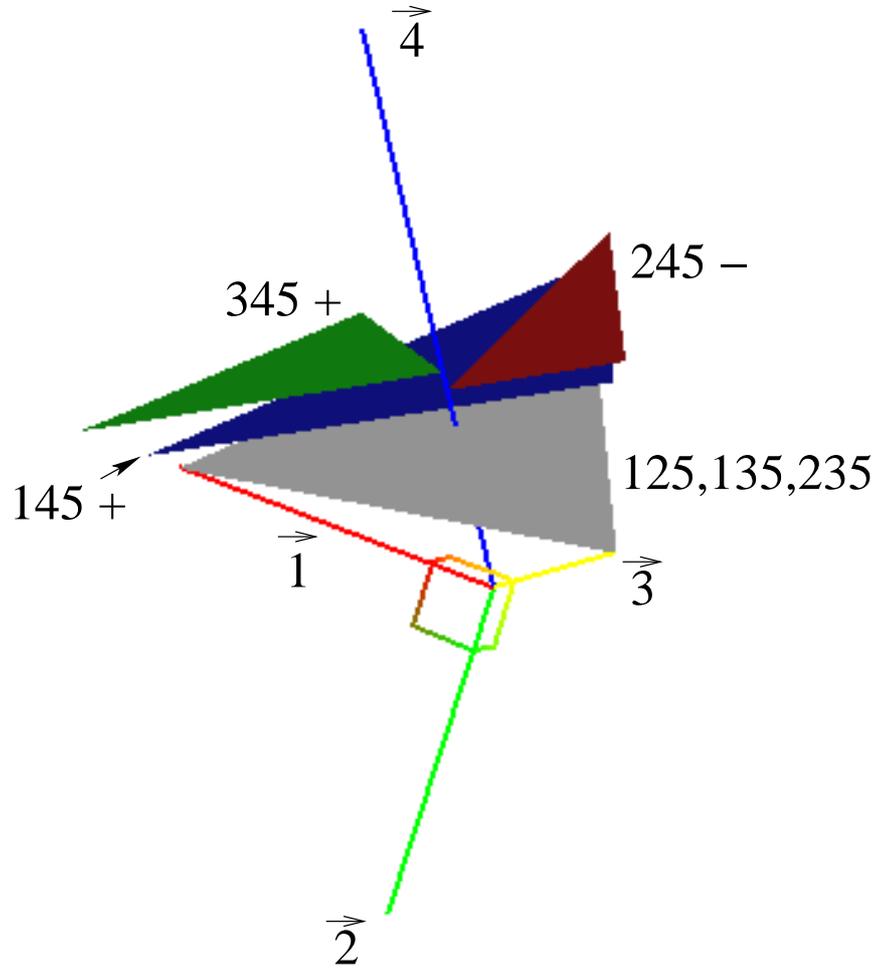}
\caption{Illustration of an unrealizable sign configuration $\vec{\epsilon}$
  for the 5-vertex,
  $\epsilon(123)=\epsilon(124)=\epsilon(125)=\epsilon(134)=\epsilon(135)=\epsilon(145)=\epsilon(234)=\epsilon(235)=-\epsilon(245)=\epsilon(345)=1$.
  We will attempt to assemble the edge tangents in such a way that these
  signs are realized for each of the ten triples of edges.
   Here the edge tangents are vectors $\vec{I}, I=1\ldots 5$.  Since
  $\epsilon(123)=1$, without loss of generality, we may place the first three
  edge tangents on the Cartesian axes: $\vec{1}$ is shown
  in red
  and lies along the $x$-axis, $\vec{2}$ is shown in green along the
  $y$-axis, and $\vec{3}$ is shown in yellow along the $z$-axis.  Now
  consider the placement of $\vec{4}$.  $\epsilon(124)=1$ requires it to lie
  in the $+z$-hemisphere, $\epsilon(134)=1$ requires it to lie in the
  $-y$-hemisphere, and $\epsilon(234)=1$ requires it to lie in the
  $+x$-hemisphere.  This leaves us with an octant, which is illustrated by a
  gray triangle, which must be pierced by $\vec{4}$.  Again without loss of
  generality, we choose $\vec{4}$ to be tangent to the vector $(1,-1,1)$.
  Now the challenge is to place $\vec{5}$ such that the six triples in which
  it is involved have the correct sign.  As for $\vec{4}$, the triples 125,
  135, and 235 require it to pierce the gray square.  $\epsilon(145)=+1$
  further requires it to be on the `+ side' of the plane defined by $\vec{1}$
  and $\vec{4}$, so it must pierce the blue triangle shown.
  $\epsilon(245)=-1$ requires it to lie on the `- side' of the
  $\vec{2}$\,$\vec{4}$ plane, so that, when restricting to a subset of the
  blue triangle, it must pierce the red triangle shown.  Finally,
  $\epsilon(345)=+1$ requires $\vec{5}$ to lie on the `+ side' of the
  $\vec{3}$\,$\vec{4}$ plane, so it must pierce the green triangle.  The fact
  that the red and green triangles do not overlap illustrates the
  unrealizability of this sign configuration.}
\label{impossible_sign_config}
\end{figure}

Because of the 
complicated structure of the recoupling computation,  
to our knowledge no one has yet attempted 
a detailed combinatorial analysis of the edge tangent information for vertices of valence greater than 4 and its effect on the volume spectrum.  
In particular it is not even obvious which triple sign configurations $\vec{\epsilon}$ can be realized at a vertex embedded in a three dimensional  
hypersurface $\Sigma$ and what $\vec{\sigma}$ combinations result from that
in the gauge invariant sector.  To give some feel for this question, we
present in figure \ref{impossible_sign_config} an example of a sign
configuration $\vec{\epsilon}$ which is not realizable in three dimensional
space.

In the present analysis we have computed the $\vec{\epsilon}$ configurations by a Monte-Carlo process, as follows.  Consider an $N$-valent vertex $v$. Now randomly sprinkle $N$ (labeled) points on the surface of a unit sphere and compute the sign $\epsilon(IJK)$ for each of the ${N \choose 3}$ triples $(IJK)$.
This gives a \signconf, which is stored in a (hash) table, along with the number of times that \signconf{} has arisen so far in the sprinkling.
At the beginning of the sprinkling process there will always be new sign configurations obtained which are successively added to the table. However, after some time,  
all the  
obtained sign configurations will already be in the table. This can be taken
as an indication that all possible sign configurations have been obtained, so
the computation is terminated.  (In particular we stop sprinkling when the
smallest number of `hits' associated to a \signconf{} is greater than one.)

A disadvantage of the sprinkling process is that one cannot be sure to have
obtained every possible sign configuration.  Furthermore, the probability of
sprinkling three points onto a single plane is zero, so the sprinkling yields only non-coplanar sign configurations.
This is the reason why in the present analysis we consider only vertices where each triple of edge tangents is linearly independent.  We will continue to call vertices with such edge configurations \emph{non-coplanar} vertices. 

Note that for non-coplanar vertices we have $\epsilon(IJK)=\pm1$ for all triples $(IJK)$  and therefore each $\sigma(IJK)$ may assume, due to (\ref{sigmas}),  the 5 values $\sigma(IJK)=0,\pm 2,\pm 4$. Thus there are in principle $5^{N-1 \choose 3}$ different $\vec{\sigma}$-configurations possible.  
The number of $\vec{\sigma}$-configurations given here is the number of numerically distinct $\vec{\sigma}$-configurations which occur when imposing gauge invariance on each of the realized $\vec{\epsilon}$-configurations.
The number of $\vec{\sigma}=\vec{0}$ configurations (i.e.\ those for which each of the ${N-1 \choose 3}$ $\sigma(IJK)$ are zero) is given by the number of occurring $\vec{\epsilon}$-configurations which lead to a $\vec{\sigma}=\vec{0}$ configuration.
The sprinkling process gives the following results:
       
\begin{table}[h] 
\center 
\begin{footnotesize}     
\begin{tabular}{|c|c|c|c|c|c|c|}
         \hline
            \cmt{1.6}{~\\valence $N$ \\ of  $v$} 
            & \cmt{1.5}{~\\number of triples: $\left(N \atop 3 \right)$\\ } 
            & \cmt{2.2}{number of theoretically possible\\$\vec{\epsilon}$-sign-configs\\ $T_{max}=2^{\left(N \atop 3 \right)}$\\[-2mm]}&   
            \cmt{2.1}{~\\number of realized $\vec{\epsilon}$-sign configs\\$~~~~~~T_R$} &
            \cmt{2}{~\\[-5mm]\center{fraction}\\[2mm] $\displaystyle\frac{T_R}{T_{max}}$\\}& 
            \cmt{1.7}{~\\number of realized\\ $\vec{\sigma}$-configs}& 
            \cmt{1.7}{~\\multiplicity $\chi_{\vec{\sigma}=\vec{0}}$ of\\ $\vec{\sigma}=\vec{0}$\\-config} 
         \\\hline\hline
         4 & 4& 16 & 16 &1 &5&6\\\hline
         5 & 10 & 1024 & 384 & 0.375 &171&24\\\hline
         6 & 20 & $2^{20}$ & 23,808 & 0.023 & 16,413&120\\\hline
         7 & 35 & $2^{35}$ & 3,486,720 &$1.015\cdot 10^{-5}$&3,079,875& 720 \\\hline
\end{tabular}
\end{footnotesize}
\caption{Sign factor combinatorics for 4--7-valent non-coplanar vertices}
\label{Sign factor combinatorics table}
\end{table}
Due to recent progress it now seems possible to overcome the mentioned disadvantages of the sprinkling process, as it is feasible to compute the possible $\vec{\epsilon}$-sign configurations directly, without using Monte-Carlo methods. The idea behind this computation is given in \cite{NumVolSpec} and will be worked out in more detail in a forthcoming paper \cite{VertexCombinatorics}. Up to now we have verified that the number of realized $\vec{\epsilon}$-sign configurations of a completely non-coplanar  5 \& 6-valent vertex is indeed given by exactly 384 and 23,808 respectively.
\\

We would like to mention here another interesting implication of the sign combinatorics given in the last column of the table above.  Equation (\ref{Volume definition gauge invariant 3}) can be interpreted as a kind of `self-regulating' property of the volume operator when applied to certain `pathological' edge configurations\footnote{
In \cite{triad I} an upper bound for the expectation value of the operator corresponding to the inverse of the classical scale factor in cosmological models is given and found to be finite. However, this bound grows with the number of edges at a vertex, and an explicit vertex configuration is given there for which this indeed happens. However, this configuration gives zero volume upon imposing gauge invariance, so it seems feasible that the problem with the explicit valence dependence of the upper bound may be cured.}
\cite{NumVolSpec}.
For vertices of valence greater than 3 there exist vertex geometries containing only non-zero edge triples, however all of them occur with identical sign, e.g.\ $\epsilon(IJK)=1$~ $\forall~ I<J<K$. Upon demanding gauge invariance such vertex geometries will have zero volume due to (\ref{Volume definition gauge invariant 3}) and (\ref{sigmas}), a statement which is \emph{independent} of the spin configuration at a vertex.\footnote{Of course the spin configuration has to allow for gauge invariance.}

\section{Spectral Analysis: Numerical Studies}
With the preparation of the previous section we now have all prerequisites for an analysis of the spectral properties of the volume operator.  While our computational implementation 
is applicable to arbitrary vertex configurations, it is limited in practice, of course, by finite computer resources. 
\subsection{Computer Resources}

We performed most computations on the Peyote high performance dual 64-bit
3.4 GHz Xeon
processor cluster, at the Max Planck Institute for Gravitational Physics in Potsdam, Germany, using
perhaps 20,000 CPU-hours in total.  Some computations were also performed on
a 2.2 GHz dual Opteron machine at Imperial College.
The eigenvalues were stored in 
188GB worth of binary data files.  

\subsubsection{Code and Speed Issues}
\label{code}
Our code is implemented within the Cactus framework (described below), and employs the LAPACK library to compute eigenvalues using singular
value decomposition, the details of which are described in
\cite{NumVolSpec}.  The set of \sigconfs{} for which we compute eigenvalues
is distributed among the processors.

Because of the smallest non-zero eigenvalues which decrease with $\jmax$, our
code is in fact bounded by the 64 bit numerical precision for the 5-vertex (and
6-vertex to a tiny extent.)  We can carry the machine computations beyond
$\jmax=\frac{44}{2}$, but the smallest non-zero eigenvalues will descend
deeper into the numerical noise.  See section \ref{results}.
At higher valences our computations were bound
in effect by disk space, because of the enormous number of \sigconfs{} at
large valence.
In order to test our code we also performed computations  on the 4-vertex. These computations were bounded simply because we use
a single byte to store the edge spins.  We could go \emph{much} further than the computations in \cite{Volume Paper}, 
if desired.

\subsubsection{The Cactus Framework}

Integration of our code into 
the Cactus high performance computing
framework \cite{cactus} affords a number of benefits, which is discussed more fully in \cite{NumVolSpec}.  In summary, the Cactus framework provides automatic parallelism, facilitates
linking to LAPACK and Fortran code, and allows a great degree of portability.
We hope that the true value of implementing our code within the Cactus
framework, however, will be that it may now 
be readily used by other groups, as a prototype for performing numerical Loop
Quantum Gravity computations, for the full theory.  
Instructions on how to download the code used for the computations described in the paper
will appear shortly.
The intent is to begin
developing a community code base for such computations, taking advantage of
the modularity and collaborative nature of Cactus to facilitate further
numerical investigations.

\subsubsection{Actual Computation}
We present in this section some results of a numerical study for 5--7-valent vertices, where we have computed the matrix representations of the volume operator for all sets of edge spins $j_1\le\ldots\le j_N=\jmax$ ($N=5,6,7$) and for all $\vec{\sigma}$-configurations according to table \ref{Sign factor combinatorics table}. As we want to treat all $\vec{\sigma}$-configurations for a given valence, we can take $\jmax$ up to a certain cutoff which is strongly limited in practice by the number of $\vec{\sigma}$-configurations.

The eigenvalues of the computed matrices are obtained using singular value decomposition, which is applicable due to the modulus in the definition (\ref{Volume definition gauge invariant 3}) of $\hat{V}$. The eigenvalues are finally collected into histograms.  The range of eigenvalues is divided into a certain number $\nbins$ of intervals. Then for each such interval, which we refer to as a histogram bin, the number of eigenvalues falling into it (occupation number) is counted.
(Note that we include the eigenvalue degeneracy in this counting, so that
each matrix $Q$ will contribute $\mathrm{dim}(Q)$ eigenvalues toward the
histograms.)
Here each such eigenvalue is regarded as coming with a redundancy
${\bf{D}}(\vec{j})\cdot \chi_{\vec{\sigma}}$ where ${\bf{D}}(\vec{j})$ is given in
(\ref{Degenracy of the spin levels}) and as described above $\chi_{\vec{\sigma}}$ denotes the number of $\vec{\epsilon}$-sign configurations giving the particular $\vec{\sigma}$-configuration.
The occupation number of the histogram bins then serves as a notion of spectral density. We can discuss the properties of this spectral density, when superposing the eigenvalues coming from different parameter ranges.  We omit the eigenvalues $\eval=0$ from all histograms, as they are much more numerous than any of the displayed occupancy numbers.
\renewcommand{\arraystretch}{1.5}
\begin{table}[htbp]
\center

\begin{footnotesize}
\begin{tabular}{|c||c|c|c|c|}\hline
case & valence $N$ & $\vec{\sigma}$-config  &$\jmax=j_N$ & $\vec{j}$-config
\\\hline\hline
1&fixed & fixed &fixed & fixed
\\\hline
2&fixed & fixed &fixed & all
\\\hline
3&fixed & fixed &all & all
\\\hline
4&fixed & all &fixed & all
\\\hline
5&fixed & all &all & all
\\\hline
6&all & all &all & all
\\\hline
\end{tabular}
\end{footnotesize}
\caption{\label{different histogram construction}Some possible choices of parameter ranges for histogram construction: `fixed' implies keeping the according parameter fixed, `all' implies placing all eigenvalues in the histogram, regardless of the value of this parameter.} 
\end{table}

Table \ref{different histogram construction} illustrates different ways to obtain histograms.
For example, \fbox{case 3} shows that for a fixed vertex valence $N$ we can collect all eigenvalues from spin configurations $\vec{j}$  
taking all values up to the spin-value cutoff $\jmax$, for fixed $\vec{\sigma}$, in order to see the effect of different vertex embeddings on the spectrum. 
We may also, as in \fbox{case 4}, combine all fixed $\vec{\sigma}$ histograms for a given vertex, but restrict to a fixed $\jmax=j_N$, to illustrate the dependence of the spectrum on $\jmax$.
Finally, as shown in \fbox{case 6}, we may take all eigenvalues we have computed into one histogram \cite{NumVolSpec}. In each histogram we plot the occupation number $\numevals$ of the bins versus the eigenvalue $\eval$.  Note that the displayed eigenvalues $\eval$ have to be multiplied by  $\sqrt{|Z|}\ell_P^3$, where due to our conventions (\ref{erste}), (\ref{Volume definition gauge invariant 3}), $Z$ is a constant and $\ell_P^3$ is the Planck length cubed.

\subsection{Results}
\label{results}
Our numerical data reveals that for vertex valence $N>4$ the sign factors contained in the definition (\ref{erste}) of the Ashtekar-Lewandowski volume operator open a fascinating interplay between spin quantum numbers and gauge invariance on the one hand, and the graph label on the other hand, which manifests itself in different sign-dependent spectral densities of this operator.
This has not been analyzed in detail before due to the already very complicated recoupling computation.
 
\begin{figure}[htbp]
\center
  \psfrag{frequency}{$\numevals$}
  \psfrag{eigenvalue}{$\eval$}
  \includegraphics[width=13.2cm]{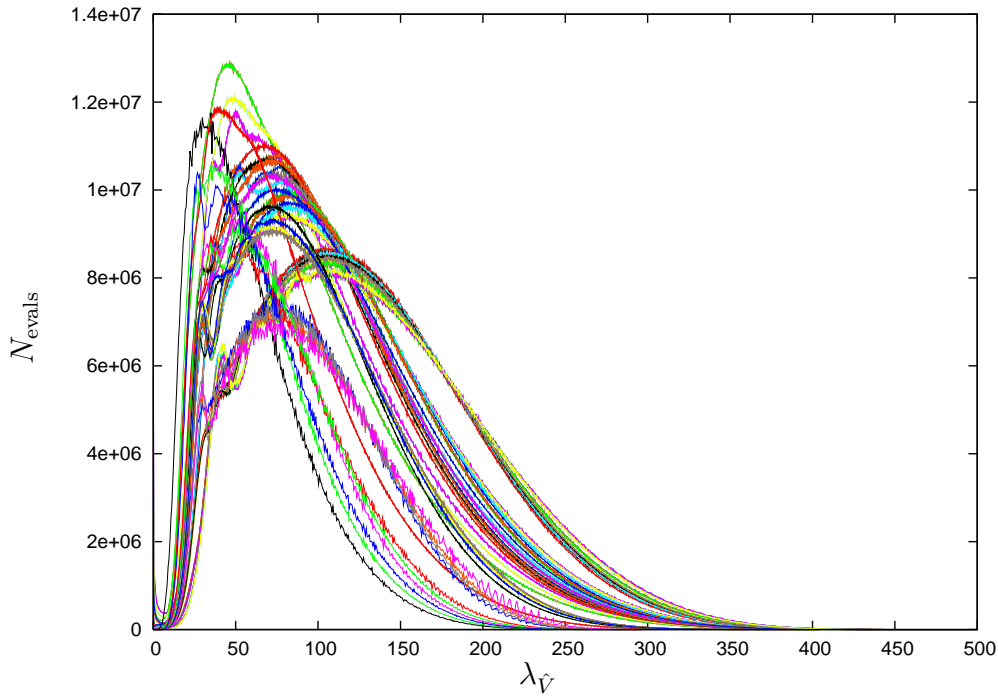}
\caption{Gauge invariant 5-vertex, case 3 of table \ref{different histogram construction}:  Histogram for each of the 85 $\vec{\sigma}$-configurations. For each fixed $\vec{\sigma}$ all eigenvalues resulting from spin configurations $\vec{j}$ up to $\jmax=\frac{44}{2}$ are collected. Note that we have treated all  
$\vec{\sigma}$-configurations equally here ($\chi_{\vec{\sigma}}\stackrel{!}{=}1$) in order to make a direct comparison possible.
All but ten of the histograms have $\nbins=2048$, the others $\nbins=512$.  These numbers are chosen to generate the sharpest curves possible.}
\label{5v_all_sigconfs}
\end{figure}

\begin{figure}[htbp]
\center
   \psfrag{number}{$\nevals$}
  \psfrag{eval}{$\eval$}
  \includegraphics[width=13.2cm]{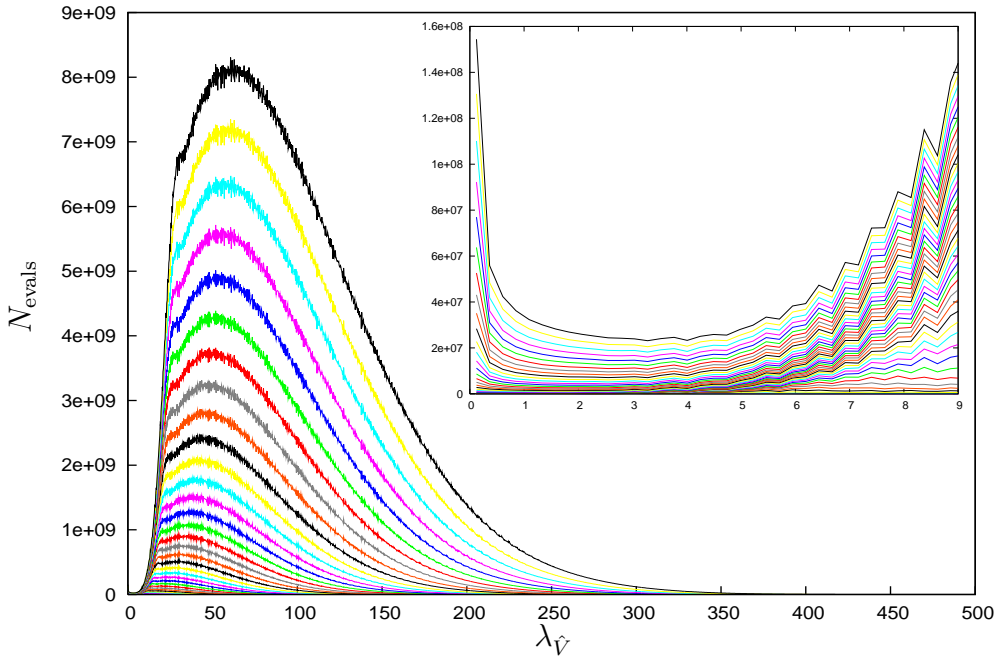}
  \caption{Gauge invariant 5-vertex, case 4 of table \ref{different histogram construction}: Histograms up to $j_{max}=\frac{44}{2}$ (top curve, below it are histograms for
   smaller $j_{max}$).  $\nbins=2048$.
   For $j_{max}=\frac{44}{2}$ there are $4.6 \times 10^{12}$ eigenvalues in all.
   Inset: Portion of histogram for $\eval\leq 9$.}
\label{Overall histogram 5-vertex}
\end{figure}

In general we find that the actual form of an eigenvalue distribution heavily depends on the sign factor combination  $\vec{\sigma}$ which is induced by the vertex geometry (\fbox{case 3} in table \ref{different histogram construction}), whereas varying $\jmax$ while taking all $\vec{\sigma}$-configurations (\fbox{case 4} in  table 
\ref{different histogram construction}) only seems to rescale the eigenvalue distribution, since a larger $\jmax$ gives more eigenvalues.
This is illustrated in figures \ref{5v_all_sigconfs} and \ref{Overall histogram 5-vertex} respectively, for the gauge invariant 5-valent vertex, which we have computed up to $\jmax=\frac{44}{2}$. As shown in table \ref{Sign factor combinatorics table} we have 171 $\vec{\sigma}$-configurations for the $5$-vertex.
Due to the modulus in the definition (\ref{Volume definition gauge invariant 3}) of $\hat{V}$ two sign configurations  $\vec{\sigma},\vec{\sigma}'$ give the same spectrum if they differ by an overall sign only, that is $\sigma(IJK)=-\sigma'(IJK)~~\forall I,J,K$. Hence (excluding $\vec{\sigma}=\vec{0}$) we effectively have 85 distinct $\vec{\sigma}$ configurations for the 5-vertex.

\begin{figure}[tbhp]
\center
\psfrag{eval}{$\mineval$}
\psfrag{sigconf}{\cmt{3}{~\\$\vec{\sigma}$-configuration}}
\psfrag{jmax}{$2\cdot \jmax$}
\includegraphics[width=14.0cm]{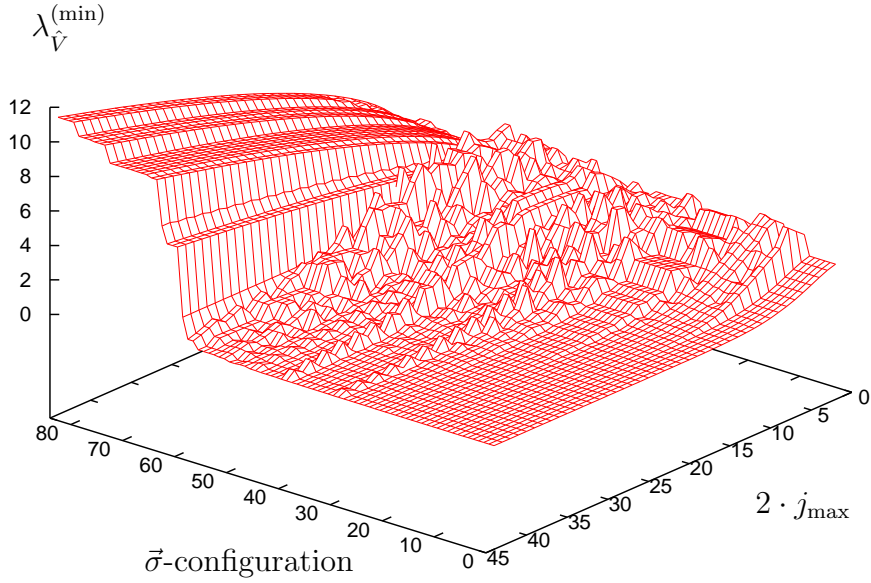}
\caption{\label{5v smallest evals}Smallest non-zero eigenvalue $\mineval$ for each of the 85 $\sigma$-configurations (arbitrarily labelled) as a function of $\jmax$, for the gauge invariant 5-vertex.  Note that multiple $\vec{\sigma}$-configurations lead to identical $\mineval(\jmax)$.  This is presented in detail in the companion paper \cite{NumVolSpec}.}
\end{figure}

Remarkably the smallest non-zero eigenvalues in a dataset can behave in three different ways as the largest spin $\jmax$ at the vertex is taken to higher values, depending on the $\vec{\sigma}$-configuration at the vertex.  They can increase (as was already known for the 4-vertex), but they can also stay constant or decrease. For the latter behavior we find sequences of smallest eigenvalues which tend to
decay towards zero exponentially with $\jmax$, for the 5- and 6-vertex.  For the 7-vertex this is difficult to decide, as we do not have enough data points, due to limited computational resources. We however find all three  behaviors of the smallest non-zero eigenvalues for the 5, 6, and 7-vertex.
This is illustrated in figure \ref{5v smallest evals}.

In the case of the histograms for the 5-vertex this results in an accumulation of eigenvalues close to zero, as can be seen in the inset plot of figure \ref{Overall histogram 5-vertex}. 
When going to larger eigenvalues the spectral density decreases to a minimum and then rises again. This second rising edge can be fitted with an exponential, which one would suppose by demanding a quasi-continuous volume spectrum at large spins.
\begin{figure}[tbhp]
\center
\psfrag{6-vertex: full spectrum up to various jmax}{}
\psfrag{frequency}{$\nevals$}
\psfrag{eigenvalue}{$\eval$}
\psfrag{jmax}{$2\cdot \jmax$}
\includegraphics[width=14.0cm]{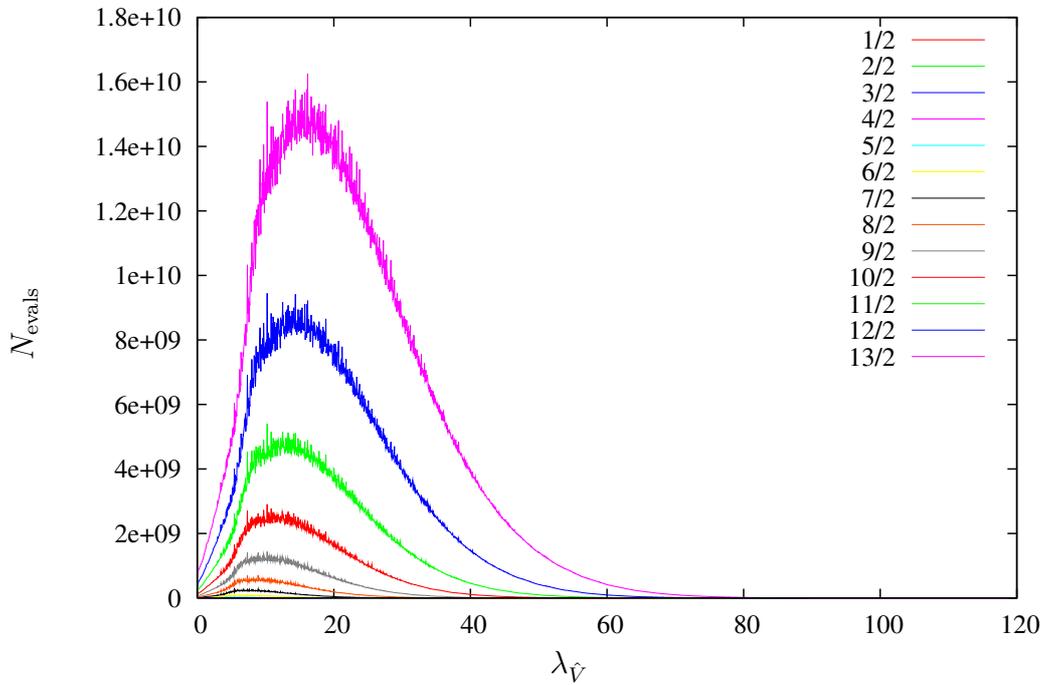}
\caption{Gauge invariant 6-vertex, case 4 of table \ref{different histogram construction}: Histograms up to $j_{max}=\frac{13}{2}$ (top curve, below it are histograms for smaller $j_{max}$).  $\nbins=2048$.
   There are $7.4 \times 10^{12}$ eigenvalues in all.}
\label{6 vertex histograms} 
\end{figure}
The gauge invariant 6-vertex, shown in figure \ref{6 vertex histograms}, also possesses the rising edge, but there is no minimum at small eigenvalues in the spectral density, the eigenvalues seem to be rather equally distributed close to zero.\footnote{
This absence of a non-zero minimum in the spectral density at the 6-vertex is
not necessarily maintained when one restricts the histogram to a single
\sigconf{}, as one can see in figure 25 of \cite{NumVolSpec}.  Thus it may be
the `averaging' over \sigconfs{} that is removing the `lip' at zero, which
may reemerge only at larger $\jmax$ (c.f.\ section 6.3.1 of \cite{NumVolSpec}).}
For the gauge invariant 7-vertex we obtain a rising edge similar to the 6-vertex, without a lip. However as can be seen in \cite{NumVolSpec}, it is not obvious if the spectral density rises exponentially. This might be caused by the fact that, due to the huge number of $\vec{\sigma}$-configurations at the 7-vertex, we have calculated the spectra only up to $\jmax=\frac{5}{2}$, so the spectrum seems to be largely dominated by the proliferation of $\vec{\sigma}$-configurations. 
Due to the necessary cutoff for $\jmax$ the rising edge in all histograms then merges into a decreasing tail (which can be considered entirely a cutoff effect), however the rising edge can be extended towards larger eigenvalues as the cutoff for $\jmax$ is increased.

\begin{figure}[tbhp]
\center
\psfrag{eval}{$\maxeval$}
\psfrag{sigconf}{$\vec{\sigma}$-configuration}
\psfrag{jmax}{$2\cdot \jmax$}
\includegraphics[width=14cm]{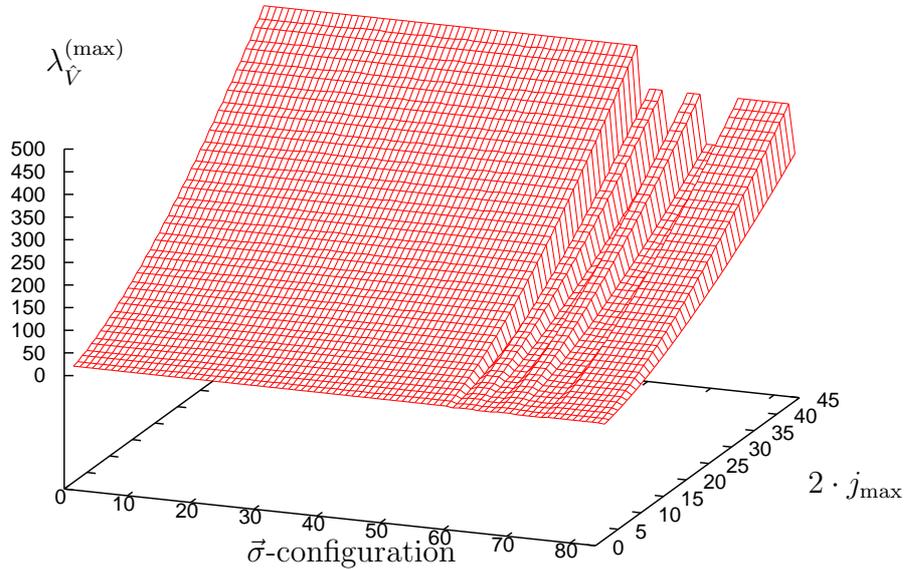}
\caption{\label{5v largest evals}Largest eigenvalues vs.\ $\jmax$ for each of the 85 $\vec\sigma$-configurations of the gauge invariant 5-vertex.  Note that many $\vec{\sigma}$-configurations lead to identical $\maxeval(\jmax)$.}
\end{figure}

The behavior of the largest eigenvalues $\maxeval$ also depends on the
$\vec{\sigma}$-configuration, as shown in figure \ref{5v largest evals} for
the 5-vertex.

Figures \ref{5v smallest evals} and \ref{5v largest evals} suggest the
presence of a non-trivial symmetry of the volume operator.  In particular,
they show groups of \sigconfs{} possessing identical smallest or largest
eigenvalues. These \sigconfs{} can be grouped into equivalence classes, which
is done explicitly in \cite{NumVolSpec}.  Unlike the $\vec\sigma \rightarrow
-\vec\sigma$ symmetry, it is not obvious why such \sigconfs{} yield identical
volume eigenvalues. One might presume that this is caused by the fact that a permutation of the edge labels at a vertex acts as a unitary transformation (see section 
\ref{The constituent matrices}) on the volume operator and in fact one has to look at the orbits of the \sigconfs{} under the permutation group \cite{VertexCombinatorics} together with a possible presence of groups of identical edge spins contained in spin configurations contributing the smallest~/~largest eigenvalues.

The cubic 6-vertex is used often, because from it one can easily generate a
graph which forms a cubic lattice.
Although it possesses coplanar edge triples, we have also analyzed it here,
and find that it belongs to the class of $\vec{\sigma}$-configurations whose smallest non-zero eigenvalue rises as $\jmax$ is increased.
\\

\begin{figure}[htbp]
  \center
  \psfrag{numevals}{$\nevals$}
  \psfrag{eigenvalue}{$\eval$}
  \psfrag{4vertex}{\hspace{-1mm}\scriptsize 4-vertex}
  \psfrag{5vertex}{\hspace{-1mm}\scriptsize 5-vertex}
  \psfrag{6vertex}{\hspace{-1mm}\scriptsize 6-vertex}
  \psfrag{7vertex}{\hspace{-1mm}\scriptsize 7-vertex}
 \includegraphics[width=15cm,height=8cm]{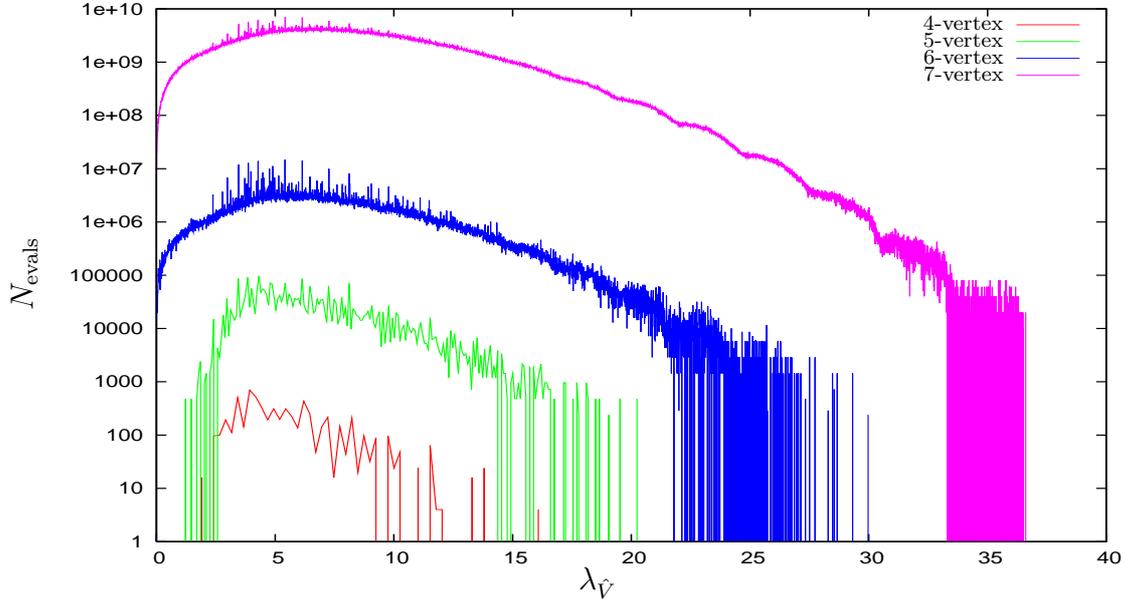}
  \caption{Collection of histograms (each constructed according to \fbox{case
  4} in table \ref{different histogram construction}) of different vertex
  valence for fixed $\jmax=\frac{5}{2}$, in log scale. This shows that higher
  valent vertices potentially contribute larger eigenvalues for the same
  maximum spin $\jmax$. The fuzzy structure at the right hand side of each
  plot is caused by a sparse population of the according histogram for the
  largest eigenvalues $\eval$ at $\jmax=\frac{5}{2}$.}
  \label{overall_hist fixed jmax} 
\end{figure}

As one can see from equation (\ref{Volume definition gauge invariant 3}), the number of  
$\sigma(IJK)$  rises with the vertex valence $N$ as ${N-1 \choose 3}$. 
Hence to compute the complete matrix representation of $\hat{V}$ we have to add the according number of constituent matrices $\hat{q}_{IJK}$. 
Figure \ref{overall_hist fixed jmax} shows a collection of plots for valences $N\!\!=\:$4--7 (each graph is constructed according to \fbox{case 4} in table \ref{different histogram construction}),  for a given maximum spin $\jmax=\frac{5}{2}$.
Note that the effect of the rising number of edge triples is obvious in figure \ref{overall_hist fixed jmax}; for example a 7-valent vertex can contribute relatively large volume eigenvalues from  
relatively small spins. This seems to be a general feature, that very high valent vertices can produce very large\footnote{Maybe even macroscopic if we go to very high valences.} volume eigenvalues, contributed by very small spins.  Such spectra are expected to be dictated by the edge combinatorics at the vertex, rather than the recoupling of spins. That is, this would give rise to `macroscopic quantum objects', as is the case when one observes quantum mechanical behavior not only for single carbon atoms but also for $C_{60}$-molecules.

We find the following characteristic spectral parameters for the different valences: The scaling of the overall (all $\vec{\sigma}$-configurations) smallest eigenvalue sequence $\mineval(\jmax)$, the overall (all $\vec{\sigma}$-configurations) largest eigenvalue sequence $\maxeval(\jmax)$, the fitted number of eigenvalues $\numevals(\eval)$ at the rising edge of the histogram for fixed $\jmax$, and the total number of eigenvalues (including redundancies)$\numevals^{(tot)}(\jmax)$ contained in a histogram.
\renewcommand{\arraystretch}{1.5}
\begin{table}[h]
\center
\cmt{15}{
\center
\begin{footnotesize}
\begin{tabular}{|l||c|c|c|c|}\hline
  \cmt{1.7}{\center valence $N$}  & $\mineval(\jmax)$& $\maxeval(\jmax)$
  &\cmt{2}{ $\numevals(\eval)$} & \cmt{2.7}{ $\numevals^{(tot)}(\jmax)$} 
  \\\hline\hline
  5&$\sim 30\cdot\mb{e}^{-0.79~\jmax}$&$\sim 3\cdot (\jmax)^{1.48}$&$2.9\cdot10^6\cdot\mb{e}^{0.43~\eval}$&$\sim 9.3\cdot 10^3 (\jmax)^{5.9}$
  \\\hline
  6&$\sim 40\cdot\mb{e}^{-2.43~\jmax}$&$\sim 3.5\cdot (\jmax)^{1.41}$&$0.19\cdot10^9\cdot\mb{e}^{0.55~\eval}$&$\sim 2.9\cdot 10^6 (\jmax)^{7.5}$
  \\\hline
  6 (cubic )&$\sim 2.1 \cdot (\jmax)^{0.49}$&$\sim 6.5\cdot (\jmax)^{1.45} $&&
  \\\hline
  7&&&$\sim 0.37\cdot 10^9\cdot\mb{e}^{0.79~\eval}$ (?)&
  \\\hline
  \end{tabular}
  \end{footnotesize}}
   \caption{Fitted parameters of the eigenvalues and spectral densities for various vertices.}
   \label{FitParameter Table}
\end{table}
\\Here we have chosen the fit functions by inspection of the data rather than from an analytic model.  For example, we can plot the smallest eigenvalue sequence $\mineval(\jmax)$ in a 
log-log plot or a log-plot and then decide whether a polynomial dependence  $\mineval(\jmax)\sim a \cdot (\jmax)^b$ or an exponential dependence  $\mineval(\jmax)\sim a\cdot \mb{e}^{b\cdot \jmax}$ gives a better fit to the data. Note however that for the rising edge of the histograms one could expect an exponentially rising number of eigenvalues from the numerical analysis of the area operator \cite{Helesfai Bene}.   
The set of data points to be used for the fitting process is chosen manually by inspection of our data (see \cite{NumVolSpec} for details). This should be kept in mind when looking at the numerical values of the fitted parameters given in table \ref{FitParameter Table}. 
Our point here is to demonstrate the possible behavior of the spectrum.
For the 7-valent vertex the fit itself is questionable, as due to the huge number of $\vec{\sigma}$-configurations our computation can only be extended to a rather small spin cutoff of $\frac{5}{2}$. Note that the growth of $\maxeval(\jmax)$ as $\sim (\jmax)^{\frac{3}{2}}$ can be understood from the actual form of the volume operator (\ref{Volume definition gauge invariant 3}), as it contains sums of $\hat{q}_{IJK}$-matrices, each coming from a polynomial of degree 3 in the spins, and finally the square root is taken.    
\\

In \cite{Meissner} a formula for the spectral density of the volume operator at the gauge invariant 4-valent vertex was derived in the limit of large spins, treating $j_1,\ldots,j_4$ as continuous classical variables.  As can be seen from  equation (\ref{gauge invariant recoupling basis for 4-vertex}) below, the recoupling process of four angular momenta $j_1,\ldots,j_4$ to a resulting total angular momentum $J=0$ implies that $j_1,j_2,j_3$ have to be recoupled to a total angular momentum $a_3\stackrel{!}{=}j_4:=j$. Now in \cite{Meissner} two different eigenvalue densities for the matrix expression 
(\ref{general matrix 4 vertex}) of the volume operator  are derived: $\rho_j(\eval)$ and $\rho(\eval)$. The former  gives an expression for the eigenvalue density  $\rho_j(\eval)|_{j_1,j_2,j_3}$ at fixed $\jmax=j_4:=j$ and fixed spins $j_1,j_2,j_3$, which is what we would need here for comparison with our spectral density.  We would have to integrate $\rho_j(\eval)|_{j_1,j_2,j_3}$ over all $j_1,j_2,j_3$ in order to get a similar situation as we have considered in our histograms, where we collect for fixed $j:=\jmax$ all eigenvalues from spin configurations $j_1\le j_2\le j_3\le \jmax$.
Unfortunately  
$\rho_j(\eval)|_{j_1,j_2,j_3}$ integrated over $j_1,j_2,j_3$ cannot be accessed explicitly, as the integral over $j_1,j_2,j_3$ involves combinations of transcendental functions of the spins. What is accessible is a so called `integrated eigenvalue density'  $\rho(\eval)_{j_1,j_2,j_3}$, which is defined as $\rho(\eval)|_{j_1,j_2,j_3}:= \int dj \rho_j(\eval)|_{j_1,j_2,j_3}$. This is equivalent to considering for a fixed set $j_1,j_2,j_3$ all possible values of $a_3=j_4$ (see section \ref{Recoupling Computation}) at once. This can be viewed as considering the spectral density of a direct sum of all matrices $Q$ of (\ref{Volume definition}) for fixed $j_1,j_2,j_3$ over all $j_4=j$, which can be reached by a recoupling of $j_1,j_2,j_3$.
While this is an interesting result, it is difficult to extract a notion of spectral density out of \cite{Meissner} as we have computed it throughout our numerical studies.  However it might be worthwhile to try applying further approximation techniques in the integrations such that properties of the total eigenvalue distribution (integrated over all spins) can be extracted.

\subsubsection*{Note on Numerical Errors}
As can be seen from table \ref{FitParameter Table} there is a serious caveat when looking at the behavior of the smallest non-zero eigenvalues.  As we go to larger $\jmax$ these eigenvalues decay exponentially to zero. However as we increase $\jmax$ the matrix entries in the matrices which are finally diagonalized become larger and larger. Hence when analyzing the smallest non-zero eigenvalues, we compute tiny numbers from relatively large numbers.  (In fact it is this effect which limits how far we can carry the computations for the 5-vertex, as mentioned in section \ref{code}.)

In order to consider the classification of $\vec{\sigma}$-configurations by the behavior of their smallest eigenvalues as a true result of our analysis it is therefore mandatory to keep track of the numerical errors in our computations. This problem turns out to be technically very puzzling. We were able to gain a good distinction between numerical errors and true results and have verified the behaviors of the smallest eigenvalues above as true properties of the volume spectrum. A detailed discussion on this issue can be found in the companion paper \cite{NumVolSpec}.

\section{\label{Analytical Results on the Gauge Invariant 4-Vertex}Analytical Results on the Gauge Invariant 4-Vertex}
In this section we will present analytical results on the spectrum of the volume operator at the gauge invariant 4-vertex which will extend some of the earlier analysis in \cite{Volume Paper}. In particular we will give a short proof that the spectrum of the volume operator at a given gauge invariant 4-vertex is simple, that is all its eigenvalues (except zero) come in pairs, and there are no further degeneracies or accumulation points in the spectrum. Furthermore we will give an explicit expression for the eigenstates of the volume operator in terms of polynomials of its matrix elements and its eigenvalues. Thirdly we will state an explicit result on the scaling of the smallest non-zero eigenvalue of the volume operator with respect to the largest spin at the vertex. This will support the numerical results of \cite{Volume Paper}. Note that the specific matrix realization of the volume operator plays a crucial role here: It can be written as an antisymmetric purely imaginary matrix having non-zero entries only on the first off-diagonal. This makes it possible to apply certain techniques from orthogonal polynomials and Jacobi-matrices  \cite{Courant Hilbert Approach Book}. This progress of course contains the drawback that in the presented results the structure of the matrix realization is heavily used. So the generalization to higher valent vertices will be technically very involved, if it is feasible at all.

\subsection{Setup}

The general case of the gauge invariant 4-vertex deals with 4 edges $e_1, \ldots,e_4$ outgoing from the vertex $v$.  Each edge $e_k$ ($k=1\ldots 4$) carries a $2j_k+1$ dimensional representation of $SU(2)$.
Due to gauge invariance the standard recoupling basis states (\ref{RC standard basis}) are given in this case as
\be\label{gauge invariant recoupling basis for 4-vertex}
   \ket{a_2(j_1\,j_2)~a_3(a_2\,j_3)\!\stackrel{!}{=}\!j_4~J(a_3\,j_4)\!=\!0~M\!=\!0}
\ee
where the intermediate recoupling $a_3(a_2\,j_3)$ has to equal $j_4$ due to the Clebsch Gordan Theorem and as a result the intermediate recoupling $a_2(j_1\,j_2)$ is the only degree of freedom for fixed spins $j_1,\ldots,j_4$.
Therefore expression (\ref{Volume definition gauge invariant 3}) for the matrix representation of the volume operator simplifies dramatically to give 
\be\label{V zw def}
  \hat{V}=\sqrt{\big|Z\cdot \sigma(123)~\hat{q}_{123} \big|}=\sqrt[4]{\big|Z|^2[\sigma(123)]^2\cdot \widehat{Q}^\dagger \widehat{Q}}
\ee
where $\widehat{Q}:=\hat{q}_{123}$ and $\sigma(123)=\epsilon(123)-\epsilon(124)+\epsilon(134)-\epsilon(234)=0,\pm 2,\pm 4$ gives a constant numerical prefactor depending on the relative orientations of the 4 edges only. In the following we will assume that $\sigma(123)\ne 0$ but leave its numerical value unspecified. Here  we set $Z=\mb{i}$. Using the explicit formula for the matrix element of the volume operator \cite{Volume Paper, NumVolSpec} we obtain for the matrix $\widehat{Q}_D$ ($D$ denotes the dimension) a matrix with the following matrix elements \cite{de Pietri,Volume Paper}:
\be\label{general matrix 4 vertex}
   \big[\widehat{Q}_D\big]_{kj}= q_k~ \delta_{k\!+\!1~j}~-~q_k~ \delta_{k~j\!-\!1} 
\ee   
where the $q_k$ are given by 
\be\label{general matrix element 4 vertex}\begin{array}{lllll}
   q_k&=&\frac{\mb{i}}{\sqrt{(2a_2+1)(2a_2-1)}}&
   \Big[(j_1+j_2+a_2+1)(-j_1+j_2+a_2)(j_1-j_2+a_2)(j_1+j_2-a_2+1)\times \\
   &&&\!\!\times(j_3+j_4+a_2+1)(-j_3+j_4+a_2)(j_3-j_4+a_2)(j_3+j_4-a_2+1)\Big]^\frac{1}{2}
\end{array}\ee
and the intermediate recoupling step $a_2(j_1\,j_2)$ is given as $a_2=a_2^{(min)}+k$,
that is
\be \label{range of a2}
\max{\big[|j_2-j_1|,|j_4-j_3|}\big]=a_2^{(min)}\le a_2 \le a_2^{(max)}=\min{\big[j_1+j_2,j_3+j_4 \big]} 
\ee
and 
$D=\dim \widehat{Q}_D =a_2^{(max)}-a_2^{(min)}+1$.

\subsection{Non-Degeneracy of the Spectrum}
By inspection of (\ref{general matrix 4 vertex}) an eigenvector
$\Psi^\lambda=(\Psi^\lambda_1,\ldots,\Psi^\lambda_D)^T$ for the eigenvalue
$\lambda$ of (\ref{general matrix 4 vertex})
with\newline$(\widehat{Q}_D-\lambda \mb{1})\Psi^\lambda=0$ will fulfill the
recurrence relation:
\be\label{Recursion Relations}\begin{array}{rrclclc}
    & q_{k-1}\Psi^\lambda_{k-1}-\lambda\Psi^\lambda_k-q_k\Psi^\lambda_{k+1}&=&0 &&\\
 \end{array}\ee
with the obvious extension for ($q_0=q_D=0$).
This recurrence relation has only one nontrivial solution $\Psi^\lambda$, for
each $\lambda\ne 0$.  (See \cite{NumVolSpec} for details.)  Thus the following
theorem holds.

\begin{Theorem}{Spectral Simplicity of $\widehat{Q}_D$} \label{Spectral Simplicity Thm}\\
  (i) The spectrum $\mbox{spec}(\widehat{Q}_D)\ne 0$ is simple (consists of $D$ real numbers).\\
  (ii) If $\Psi^\lambda=(\Psi^\lambda_1,\ldots,\Psi^\lambda_D)^T$ is an eigenvector of $\widehat{Q}_D$ ($\widehat{Q}_D\Psi^\lambda=\lambda\Psi^\lambda$, $\lambda\ne 0$) then $\Psi^\lambda_1\ne 0, \Psi^\lambda_D\ne 0$.
  
\end{Theorem}

\subsection{Eigenstates in Terms of Eigenvalues}
Starting from ($k=1$) in (\ref{Recursion Relations}) by setting $\Psi_1^\lambda=x=\mathrm{const}$ one can explicitly construct the components $\Psi^\lambda_k$. One gets\footnote{Note that (\ref{Eigenvectors 4 vertex}) can be compared to results of \cite{Carbone Carfora Marzouli} which where communicated to the authors during the preparation of this paper. } (using the integer number $L\ge 1$ with $L \le \frac{D}{2}$ for even $D$ and $L \le \frac{D-1}{2}$ for odd $D$):
\begin{eqnarray}\label{Eigenvectors 4 vertex}
\footnotesize 
   \Psi^\lambda_{2L}=\frac{-x\lambda}{\prod\limits_{k=1}^{2L-1}q_k}
   \left[\lambda^{2(L-1)}+\sum_{M=1}^{L-1}
            b_{M}^{(2L)}
             \cdot
            \lambda^{2(L-1-M)} \hspace{-2mm}
   ~~\right]
   ~~~~~
   \Psi^\lambda_{2L+1}=\frac{x}{\prod\limits_{k=1}^{2L}q_k}
    \left[\lambda^{2L~~~}+\sum_{M=1}^{L}
             b_{M}^{(2L+1)}
             \cdot
            \lambda^{2(L-M)~~~} \hspace{-2mm}
   \right]
   \nonumber
\end{eqnarray}
where we have defined
\be\label{char poly coeff}
   b_{M}^{(k_0)}:= \sum_{k_M=2M-1}^{k_0-2}~
	    \sum_{k_{M-1}=2(M-1)-1}^{k_M-2}\hspace{-2mm}
	    \ldots
	    \sum_{k_l=2(M-l)-1}^{k_{l-1}-2} \hspace{-2mm}
	    \ldots  
	    \sum_{k_2=3}^{k_3-2}
	    \sum_{k_1=1}^{k_2-2} 
	    q_{k_M}^2q_{k_{M-1}}^2\ldots q_{k_l}^2\ldots q_{k_2}^2 q_{k_1}^2
\ee
Note that if $M=1$ in this sum one has to take the upper bound $k_0-2$ of $k_M$.
One may explicitly check that these states fulfill (\ref{Recursion Relations}). 
 In order to fulfill (\ref{Recursion Relations}) for ($k=D$),
$\lambda$ has to be an eigenvalue, that is a root of the (real) characteristic polynomial
\ba \label{char poly}
\begin{array}{lclllll}
   \pi_{2L}(\lambda)&=&\displaystyle\lambda^{2L}+\sum_{M=1}^L b_M^{(k_0)}\lambda^{2(L-M)} &~~~~~~&D=2L~~\mbox{even}&~~~~~&k_0=2L+1=D+1  
   \\
   \pi_{2L+1}(\lambda)&=&\displaystyle\lambda\Big[\lambda^{2L}+\sum_{M=1}^L b_M^{(k_0)}\lambda^{2(L-M)}\Big] &~~~~~~&D=2L+1~~\mbox{odd}&~~~~~&k_0=2L+2=D+1~~~~~~~~
\end{array}
\ea
with coefficients $b_M^{(k_0)}$ defined as in (\ref{char poly coeff}).
Since the  
square-bracketed terms in the characteristic polynomials (\ref{char poly}) contain only even powers of $\lambda$ we may replace $\lambda^{2k}=\Lambda^k$, with $\Lambda=|\lambda|^2$ in order to arrive at a reduced purely real notation for them:
\be
   0=\lambda^{2L}+\sum_{M=1}^L b_M^{(k_0)}\lambda^{2(L-M)} 
   =\Lambda^L + \sum_{M=1}^L (-1)^{M}~\big|b^{(k_0)}_M\big|~\Lambda^{L-M}
   \nonumber
\ee
where we have pulled out a prefactor $(-1)^M$ coming from the purely imaginary nature of the matrix elements (\ref{general matrix element 4 vertex}). This confirms the results obtained in
\cite{Carbone Carfora Marzouli}.

\subsection{Lower Bound on the Spectrum}

Under fairly general assumptions on the scaling of the smallest three spins with respect to the fourth largest spin at a vertex, an explicit lower bound for the smallest non-zero eigenvalue can be obtained analytically, as well as the scaling of this lower bound with respect to the maximal spin. The idea is now  
to use Gershgorin's theorem \cite{Marcus Minc}. This usually provides an upper bound for the modulus of the maximum eigenvalue for a general matrix in terms of the maximal row/column sums of the modulus of matrix elements. 

However, an upper bound for the eigenvalues of the inverse matrix
$\widehat{Q}_D^{-1}$ certainly provides a lower bound for the smallest
non-zero eigenvalue of $\widehat{Q}_D$. Although this idea was presented
already in \cite{Volume Paper}, the issue was left open there, due to the
occurrence of a zero contained as a single eigenvalue in the spectrum of
$\widehat{Q}_D$ in the case of odd dimension $D$, preventing a straight
forward inversion.
 Using certain properties of the resolvent of $\widehat{Q}_D$ \cite{Courant Hilbert Approach Book}, it is possible to overcome this problem.

For any $D$ the spectrum of $\widehat{Q}_D$ can be related to the spectrum of the $(D-1)$-dimensional matrix $\WT{\widehat{Q}}_{D-1}$ which is obtained by deleting the last row and column of the matrix $\widehat{Q}_D$ in (\ref{general matrix 4 vertex}). As can be shown \cite{NumVolSpec} the spectra of $\widehat{Q}_D$ and $\WT{\widehat{Q}}_{D-1}$ interlace, that is between any two eigenvalues of $\widehat{Q}_D$ there is exactly one eigenvalue of $\WT{\widehat{Q}}_{D-1}$. In case of odd dimension $D$, $\WT{\widehat{Q}}_{D-1}$ is even dimensional and can be inverted without any problems, thanks to the fact that even dimensional matrices of the form (\ref{general matrix 4 vertex}) with non-vanishing entries on the first off-diagonal have no zero in their spectrum \cite{Volume Paper}. After an admittedly lengthy derivation presented in the companion paper \cite{NumVolSpec}, a lower bound for the smallest non-zero eigenvalue $\lambda_{\hat{V},\mbox{4-vertex}}^{(min)}$  of (\ref{V zw def}) in leading order of the maximal spin $j_{max}$ is obtained as
\be
   \lambda_{\hat{V},\mbox{4-vertex}}^{(min)}(\jmax)\ge \ell_P^3\sqrt{|Z|\cdot|\sigma(123)|\cdot j_{max}}  
\ee
Here $\sigma(123)=0,\pm 2,\pm 4$ is defined as in 
(\ref{Volume definition gauge invariant 3}), (\ref{V zw def}).
This may be compared to the numerical result \linebreak $\lambda_{min}= \ell_P^3\cdot   \sqrt{\big|Z\cdot \sigma(123) \big|\cdot2\cdot\sqrt{ j_{max}(j_{max}+1)}}$ obtained in \cite{Volume Paper} contributed by the spin configuration $j_1=j_2=\frac{1}{2}$ and $j_3=j_4=j_{max}$.
Remarkably the estimate presented in \cite{NumVolSpec} even explains
this combination of spins, that is in the upper bound approximation this particular combination of spins is enforced in order to maximize the row/column sum of the inverse matrix and can thus be understood analytically.

\section{Summary and Outlook}
In this paper
we present the results of an analysis of the operator corresponding to the classical expression for the volume of a spatial region (as constructed by Ashtekar and Lewandowski) within Loop Quantum Gravity. 
As the action of the volume operator can be decomposed into the action on single vertices, we can restrict our attention to a single vertex only.  
Consequently we have studied the volume operator at gauge invariant 4,5,6,7-valent vertices.
The companion paper \cite{NumVolSpec} contains further details.

We have started from \cite{Volume Paper}, where the matrix elements of the volume operator were explicitly derived within a framework using recoupling theory of angular momenta.  This results in a drastically simplified matrix expression, which is a crucial ingredient and starting point for the work presented here.

For the gauge invariant 4-vertex we obtain consistency with the results obtained in \cite{Volume Paper}. 
Moreover we find an analytic derivation for a lower bound of the non-zero eigenvalues and their scaling behavior with respect to the maximum spin incident at the vertex. We are able to give a simple proof for the spectral simplicity of the single matrix expressions of the volume operator at the 4-vertex. We also find an expression for the eigenstates in terms of the matrix elements and the eigenvalues of the volume operator compatible with \cite{Carbone Carfora Marzouli}.
In order to achieve these results we heavily take advantage of the band-structure of the matrix, which enables us to use certain techniques for Jacobi-matrices \cite{Courant Hilbert Approach Book}. However, this implies that it will be difficult  
to extend these analytical investigations to the general matrices arising from higher valent vertices.
Consequently a numerical study appears to be the viable way to analyze the properties of the volume operator at valences greater than 4.

There are numerous technical difficulties to be overcome in order to do numerical computations for higher valent vertices.  We have shown how the gauge invariant basis states can be implemented on a computer using the rules for the recoupling of angular momenta. A more difficult task is then to understand the sign-factor combinatorics of possible embeddings of the edges of the underlying graph at a vertex, that is the possible relative orientations of the edge tangents at the vertex. We have used a Poisson sprinkling process in the actual computation, excluding vertices with coplanar edges. However we have also pointed towards an analytical\footnote{Of course, the solutions of the inequality systems as presented in \cite{NumVolSpec} use a computer algebra program. However the actual computations will be exact  
in the sense that there are no numerical errors present when working with these solution tables of systems of inequalities.}
solution of this problem in terms of coupled systems of inequalities.
 We could then numerically compute the matrix representations of the volume operator for all sets of edge spins $j_1\le\ldots\le j_N=\jmax$ ($N=5,6,7$). Here we can take $\jmax$ up to a certain cutoff which is determined by the limitations of 
 the computational resources. The eigenvalues of these matrices are computed using singular value decomposition. The eigenvalues are finally collected into histograms,
whose bin occupation numbers serve
as a notion of spectral density. We finally discuss the properties of this spectral density, where we superpose the eigenvalues coming from all spin configurations and/or sign-factor combinations.      

 Our numerical results reveal that the sign factors contained in the definition (\ref{erste}) of the Ashtekar-Lewandowski volume operator open a fascinating interplay between spin quantum numbers and gauge invariance on one side and the graph geometry on the other side.
This manifests itself in the fact that the spectral density of the volume operator depends on the sign configuration at a vertex, which is induced by the vertex geometry.
This aspect has not been analyzed in detail before due to the already very complicated recoupling computation.  
\\

Having obtained these results the following tasks and questions occur as
important for future investigation.
     
We have seen that the volume operator is sensitive to the vertex geometry as manifested in the relative orientation of edge-tangents at a vertex which result in a particular spatial diffeomorphism invariant sign-factor combination. 
Here it is important to understand what kind embeddings of $N$ edges in three dimensional Riemannian space, labelled by sign factors $\vec{\epsilon}$, can be realized at all. Here we also have to understand how to factor out permutations of the edges. These issues will be worked out in detail in a forthcoming publication \cite{VertexCombinatorics}.  
It will be interesting to see how the number of realized embeddings of $N$ edges grows as $N$ is increased, in particular if there is a kind of saturation process of that number as we send $N$ to infinity. (e.g.\ there is the possibility that the growth in the number of possible embeddings rapidly decreases as the number of edges is increased.)
 This could also provide us with an answer to the question of whether it is justified to restrict oneself  
to graphs which have vertices only up to a certain maximum valence. 

In the analysis above, it appears that e.g.\ a 7-valent vertex can contribute larger volume eigenvalues, than lower valent vertices for a given maximum spin, as can be seen from figure \ref{overall_hist fixed jmax}. Could it be a general feature that very high valent vertices produce very large (even macroscopic) volume eigenvalues even from small spins, or is there some mechanism which suppresses such contributions from large valence vertices?

Understanding the sign-factor combinatorics and their encoded geometrical information is important for a number of reasons.   
Eventually one would like to  
find a combinatorial formulation of Loop Quantum Gravity \cite{Bombelli: Voronoi Complexes}.  Are the sign factors related 
to local geometrical quantities of cell-decompositions of the three dimensional spatial foliation surfaces?   

Another consideration is that it is  
desirable to gain more control, that is a geometric interpretation, on the quantum numbers of states in the (spatial) diffeomorphism invariant Hilbert space \cite{TT:big script} of the theory.  One seeks states which are, roughly speaking, constructed by an averaging of a graph over all its possible (diffeomorphic)
embeddings into the spatial foliation hypersurfaces. 
These diffeomorphism invariant states could be classified (in addition to the moduli parameters  of \cite{Rovelli: Moduli Spaces}) by the volume spectrum\footnote{Note that the region whose volume is measured also has to be specified in a diffeomorphism invariant manner, for instance by a matter field.} caused by the sign 
configurations at the vertices which are induced by the diffeomorphism invariant structure of the graph.
~     
\\

Given the fact that we now have at hand a  
numerical treatment of the volume operator in full Loop Quantum Gravity,
it will be worthwhile to look again at the suggestion pointed out in \cite{TT: GCS I},  to use the volume operator in the construction of complexifier coherent states rather than the area operator.  This
may remove certain embedding ambiguities from which the current complexifier coherent states suffer \cite{Sahlmann: CS for CQG}, which  
result from the usage of the area operator as complexifier.\footnote{A surface has to be chosen which is punctured by the underlying graph, giving rise to eigenvalues of the area operator. This surface can be tilted with respect to the graph, such that the number of transversely intersecting edges changes, 
which makes  
the complexifier eigenvalues a matter of choice.} There are certain conceptual issues still to be solved, however,
e.g.\ that due to the huge kernel of the volume operator these coherent states are not normalizable in the construction proposed so far.
Nevertheless this can be a  
promising starting point  
to (at least numerically) construct (spatial) diffeomorphism invariant coherent states for concrete computations, 
using the sign factors as  labels  
to denote the peakedness of the thus constructed states around a given (classical) geometry.  Such a notation has not been available up to now, which has prevented the construction of these states. 
\\

We would like to mention here that the question of what category of diffeomorphisms and edges should be considered as admissible for LQG is a lively research area. In  \cite{Sawin Baez} the use of the smooth category is discussed, while \cite{Zapata:1997db} looks at the possibility of using the piecewise linear category. In \cite{Fairbairn:2004qe} diffeomorphisms which are analytic everywhere except for a finite number of points are analyzed. Finally, as pointed out in \cite{Fleischhack:2004jc} and applied in \cite{Lewandowski:2005jk},  it appears that one has to stick to the stratified analytic category, which roughly allows that an edge can get a finite number of kinks under such transformations. See also \cite{Koslowski:2006nr} on this subject. Of course, the question of what information should enter diffeomorphism invariant states has to be considered within this context, and it could for instance be that one has to use states which are additionally averaged over all $\vec{\sigma}$-configurations, and thus do only contain information on the linking structure of the underlying graph\footnote{From this perspective it might be interesting to look at the question of whether the operator  $\hat{V}^{(RS)}$ of \cite{Rovelli Smolin Volume Operator} can be related to the averaged $\hat{V}^{(AL)}$ analyzed in the present paper.}.
\\

Another potentially important direction we would like to point at is the possible application of  techniques from  random matrix theory \cite{Mehta: Random Matrices} to Loop Quantum Gravity.
By inspection of the volume spectrum averaged over  
the $\vec{\sigma}$-configurations  and spins as in figure \ref {Overall histogram 5-vertex}, we can see a certain similarity of the produced spectral densities to the densities occurring in the context of certain ensembles of matrices having entries which are independent random variables. The idea here is the following.  The averaging of the volume spectrum over all spin configurations $j_1,\ldots, j_{N}$ of the $N$-valent vertex $v$  (up to a certain maximum spin $j_{max}$ ) can equivalently be seen as probing a set of random numbers $j_1,\ldots, j_{N}$  all occurring with the same probability. The problem to be solved in this direction is to extend the notion of random matrices to matrices whose entries are not random variables themselves but functions of a set of random variables, that is the spins $j_1,\ldots, j_{N}$.
For such an analytical treatment it will be necessary to further investigate universal properties of the spectral density of the volume operator.\footnote{That is an asymptotic (with respect to $\jmax\rightarrow\infty$) shape of the `normalized' (rescaled such that the largest eigenvalue and the largest occupation number are set to 1) eigenvalue distribution as well as the components of the eigenstates.}
\\

Finally we would like to point out that $\hat{V}$ is not only essential in the construction of the operator  corresponding to the Hamilton constraint \cite{TT:QSD} (as shown in (\ref{Hamilton constraint})),
but in all kinds of matter Hamiltonians \cite{TT:QSD V} in the theory. Having achieved a computational treatment of the volume operator now opens the door to numerically test the results of symmetry reduced models \cite{Bojowald:2006da,Ashtekar:2006es,Husain:2003ry} within full Loop Quantum Gravity. 
As the code of the present analysis will be publicly released, we hope that this might serve as a starting point to developing a computational toolkit for numerical Loop Quantum Gravity, which is freely available to the research community.
\\

Concluding, we would like to state that despite the conceptual problems which are inherent to Loop Quantum Gravity, it is important to 
study the physical implications of the theory in its current form.
Numerical analysis now opens the door to new insights into the predictions of Loop Quantum Gravity, as it allows the study of aspects of the theory which are inaccessible using analytical techniques.  
The insights thus obtained provide important feedback to the development of the theory.
To facilitate future numerical analysis,  
we intend to release the source code we have used in this investigation, 
as `thorns' for the Cactus high performance computing framework.
We hope this will a first step in the development of a community code base for a wide range of computations in Loop Quantum Gravity.

\section*{Acknowledgments}
We would like to thank Thomas Thiemann for encouraging us to start this project and for suggestions and discussion on the issue of the sign factors.
His continuous, reliable support for this project and his helpful suggestions on the manuscript are gratefully acknowledged. 
We thank Steve White for pointing out to us the relative merit of singular
value decomposition, because of its numerical stability.  This resulted in a
significant improvement in the quality of our numerical results.
Moreover we thank Jonathan Thornburg for providing  
perspective into the problem of handling numerical error.

We would like to thank Luciano Rezzolla and the numerical relativity group
at the Albert Einstein Institute Potsdam for support and use of computational resources at the AEI, in particular the Peyote cluster.
In addition we thank the Perimeter Institute for Theoretical Physics, where parts of this project were completed, for hospitality.

JB thanks the Gottlieb-Daimler and Carl-Benz foundation, the German Academic Exchange Office as well as the Albert Einstein Institute for financial support. Furthermore JB has been supported in part by the Emmy-Noether-Programme of the Deutsche Forschungsgemeinschaft under grant FL 622/1-1. 
The work of DR was supported in part through the Marie Curie Research and
Training Network ENRAGE (MRTN-CT-2004-005616).

We thank Simone Speziale for pointing out \cite{Carbone Carfora Marzouli} to us.

We are grateful for valuable suggestions on improving the manuscript by
anonymous referees of Classical and Quantum Gravity.

\end{document}